\documentclass[aps,prd,onecolumn,groupedaddress,showpacs,nofootinbib,amssymb]{revtex4}
\usepackage{amssymb}
\usepackage{amsmath}
\usepackage{tensor}
\usepackage{amsfonts}
\usepackage{bm}
\usepackage{cancel}
\usepackage{comment}
\usepackage{filecontents}
\usepackage{graphicx}
\usepackage{color}
\usepackage[lofdepth,lotdepth]{subfig}
\usepackage{floatrow}

\allowdisplaybreaks[4]

\begin{document}

\title{Autonomous Dynamical System of Einstein-Gauss-Bonnet Cosmologies}
\author{N. Chatzarakis,$^{1}$\, \thanks{nchatzar@physics.auth.gr} V.K. Oikonomou,$^{1,2,3,4}$\thanks{v.k.oikonomou1979@gmail.com}}
\affiliation{
$^{1)}$ Department of Physics, Aristotle University of Thessaloniki, Thessaloniki 54124, Greece\\
$^{2)}$ Laboratory for Theoretical Cosmology, Tomsk State
University of Control Systems
and Radioelectronics, 634050 Tomsk, Russia (TUSUR)\\
$^{3)}$ Tomsk State Pedagogical University, 634061 Tomsk, Russia\\
$^{4)}$Theoretical Astrophysics, IAAT, University of T\"{u}bingen,
Germany }

\tolerance=5000

\begin{abstract}
In this paper, we study the phase space of cosmological models in
the context of Einstein-Gauss-Bonnet theory. More specifically, we
consider a generalized dynamical system that encapsulates the main
features of the theory and for the cases that this is rendered
autonomous, we analyze its equilibrium points and stable and
unstable manifolds corresponding to several distinct cosmological
evolutions. As we demonstrate, the phase space is quite rich and
contains invariant structures, which dictate the conditions under
which the theory may be valid and viable in describing the
evolution Universe during different phases. It is proved that a
stable equilibrium point and two invariant manifolds leading to
the fixed point, have both physical meaning and restrict the
physical aspects of such a rich in structure modified theory of
gravity. More important we prove the existence of a heteroclinic
orbit which drives the evolution of the system to a stable fixed
point. However, while on the fixed point the Friedman constraint
corresponding to a flat Universe is satisfied, the points
belonging to the heteroclinic orbit do not satisfy the Friedman
constraint. We discuss the origin of this intriguing issue in some
detail.
\end{abstract}


\maketitle

\section{Introduction}

The idea of extending General Relativity is a necessity imposed by
the inadequacy of General Relativity to sufficiently explain
issues such as the initial singularity, the early stages of the
evolution of the Universe and also for example the late-time
acceleration of the Universe. It was in fact proved that slight
modifications in the Einstein-Hilbert action, such as the addition
of quadratic curvature terms, can in fact resolve some of these
issues, or can alternatively describe phenomena, without the need
of additional scalar fields. At the same time, the necessity to
cope up with a quantum foundation of gravity, led to the
perception of these quadratic terms as second-order corrections of
General Relativity attributed to either a -yet- unknown quantum
theory of gravity, or to a string theory, given of course the fact
that General Relativity is a linearized version of such a
higher-order theory, or the low-energy limit of some string
theory.

A few years prior to that, Lovelock proposed a Lagrangian
formulation of $n$-dimensional gravity, that would generalize
Einstein's General Relativity by adding higher-order terms of
curvature \cite{Lovelock:1971yv, Farhoudi:1995rc}. In this
generalized theory, General Relativity is merely the first-order
approximation, that cannot describe cases where strong gravity is
implied, such as the very early Universe or the interior of a
black hole. In this stream, one may consider the second-order
approximation of the theory, that is, the inclusion of both the
Ricci scalar and the Gauss-Bonnet invariant. The former realizes
General Relativity as we know it, while the latter, being a
topology-related term in $4$-dimensional spacetimes, and a trivial
term in all higher-than-four dimensions, may sufficiently describe
the strong gravity effects. Models that combine scalar fields with
the Gauss-Bonnet invariant are called the Einstein-Gauss-Bonnet
models and can be further generalized to have the form
$f(\mathcal{G})$ and $f(R,\mathcal{G})$, where arbitrary functions
of the Ricci scalar and of the Gauss-Bonnet term are considered
instead of them. For a brief introduction to either
$f(\mathcal{G})$ or Einstein-Gauss-Bonnet, the reader is referred
to the reviews
\cite{reviews1,reviews2,reviews3,reviews4,reviews5,reviews6} and
for several alternative theories of modified gravity.

Initially, questions regarding the viability of
Einstein-Gauss-Bonnet theories of gravity have been discussed in
\cite{Chingangbam:2007yt}. Consequently, many would regard both as
theories not capable of dealing with the actual Universe. However,
many cosmological models based on the $f(\mathcal{G})$ or the
$f(R,\mathcal{G})$ models have been developed so far -see Refs.
\cite{Nojiri:2005am, Sanyal:2006wi, Cognola:2006eg, Li:2007jm} for
some early examples, and \cite{Escofet:2015gpa, Odintsov:2016hgc,
Oikonomou:2016rrv, Oikonomou:2017ppp, vandeBruck:2017voa,
Bamba:2017cjr, Fomin:2017vae, Fomin:2017qta, Houndjo:2017jsj,
Saridakis:2017rdo} for more recent considerations. In most of the
cases, the Gauss-Bonnet term can generate acceleration in the
Universe, just like the Cosmological Constant and hence, these
theories can harbor phenomena in the Universe such as the observed
late-time acceleration. In the same manner, a vivid discussion
about the stability of inflationary scenarios \cite{Li:2007jm,
Bamba:2009uf, vandeBruck:2016xvt, Oikonomou:2017ppp,
Santillan:2017nik, Houndjo:2017jsj} or bounce cosmological models
\cite{Bamba:2014zoa, Escofet:2015gpa, Mathew:2018rzn} appeared in
the last decade. Also in Ref. \cite{Makarenko:2012gm} exact
solutions are presented for many kinds of singularities in plain
Gauss-Bonnet gravity. Many of the above works contain not only the
Gauss-Bonnet invariant, but also a scalar field non-minimally
coupled to it, while Refs. \cite{Kanti:2015pda, Kanti:2015dra}
claim that the Einstein terms can be ignored and the scalar field
and the Gauss-Bonnet term can dominate at early times. Also in
Ref. \cite{vandeBruck:2017voa} it is shown that the Gauss-Bonnet
term is negligible and the inflation era is mainly generated by
the potential of the scalar field, pretty much like the more
traditional approaches.

Vital support to these theoretical approaches came when successful
compactifications from $1+3+D$- to $1+3$-dimensional spacetimes
were proved possible \cite{Canfora:2016umq, Toporensky:2018xpo,
Pavluchenko:2018cmw}. Similar results concerning the linear and
non-linear dynamics of the theory, for either astrophysical or
cosmological solutions can be found in the literature, for example
see Refs. \cite{Shinkai:2017xkx, Toporensky:2018xpo} for a
numerical approach and \cite{Fomin:2017vae, Fomin:2017qta} for an
analytical approach. As a result, any higher-dimensional Lovelock
or Einstein-Gauss-Bonnet theory (see for example cosmological
models of Refs. \cite{Pavluchenko:2016wvi, Pavluchenko:2016hfu,
Pavluchenko:2017svq, Pavluchenko:2018sga} or
\cite{Ivashchuk:2009hi, Ivashchuk:2016jpe}) can be dynamically
compactified to a $1+3$ Friedmann-Robertson-Walker (FRW) Universe,
described by a scalar-Einstein-Gauss-Bonnet theory. In principle,
a string-inspired theory can be associated with a classical
modified gravity in the form of a scalar field and a quadratic
curvature term, however it is also possible that an infinite tower
of higher-curvature terms (besides just quadratic terms) may be
present, and there could in principle be additional fields at low
energies besides a single scalar. In the literature, string theory
motivated modified gravities have produced viable results,
compatible with the observational data \cite{Bamba:2007ef,
Glavan:2019inb, Odintsov:2018zhw, Nojiri:2019dwl}, yet several
questions remain unanswered. Among the cosmological issues
covered, we could name those of bounce cosmologies
\cite{Koshelev:2013lfm, Oikonomou:2015qha}, or inflationary
scenarios \cite{Chakraborty:2018scm, Odintsov:2018zhw,
Nojiri:2019dwl}. Astrophysical issues have also been examined,
such as the spherical collapse of matter \cite{Maeda:2006pm,
Benkel:2016rlz, Abbas:2018ica}.

In the present paper we shall study the phase space of the
Einstein-Gauss-Bonnet cosmological models, focusing on the
viability and stability issues of the theory, so rigorously
discussed in the literature. We shall investigate whether the
dynamical system of the theory can be an autonomous dynamical
system, one that exposes actual attractors or repellers in the
form of equilibrium points. In such a dynamical system, we shall
study the invariant substructures of the phase space, namely the
equilibrium points, stable and unstable manifolds and so on. These
phase space structures provide vital information about the
dynamical implications of the theory, only if the dynamical system
is autonomous. In the literature, the autonomous dynamical system
approach is quite frequently adopted
\cite{Oikonomou:2019muq,Oikonomou:2019nmm,Odintsov:2018uaw,Odintsov:2018awm,Boehmer:2014vea,Bohmer:2010re,Goheer:2007wu,Leon:2014yua,Guo:2013swa,Leon:2010pu,deSouza:2007zpn,Giacomini:2017yuk,Kofinas:2014aka,Leon:2012mt,Gonzalez:2006cj,Alho:2016gzi,Biswas:2015cva,Muller:2014qja,Mirza:2014nfa,Rippl:1995bg,Ivanov:2011vy,Khurshudyan:2016qox,Boko:2016mwr,Odintsov:2017icc,Granda:2017dlx,Landim:2016gpz,Landim:2015uda,Landim:2016dxh,Bari:2018edl,Chakraborty:2018bxh,Ganiou:2018dta,Shah:2018qkh,Oikonomou:2017ppp,Odintsov:2017tbc,Dutta:2017fjw,Odintsov:2015wwp,Kleidis:2018cdx,Oikonomou:2019boy}.
We shall consider several cosmological scenarios, such as the de
Sitter expansion, as well as matter and radiation domination
evolutions. The most important outcome of our analysis is the
existence of a heteroclinic orbit which drives the dynamical
system to the stable physical fixed points. However, an intriguing
phenomenon we discovered is that the points along this curve do
not always satisfy the Friedman constraint for the flat FRW
geometry we chose, if the matter terms (dust and radiation) are
not take into account. This originally seems to be contradict the
well-established theory, that a cosmological model cannot evolve
from a non-flat geometry to a flat attractor, however such a
perception falls victim to a vacuum Universe assumption -one that
we do not follow.

The paper is organized as follows: In section II, we present the
theoretical framework of the Einstein-Gauss-Bonnet theory, derive
the field equations and conservation laws and specify them for a
FRW Universe. In section III, we construct the dynamical system
corresponding to the Einstein-Gauss-Bonnet theory, and we
investigate the conditions under which it can be autonomous; as it
is proved, the cases for which the dynamical system is rendered
autonomous correspond to distinct cosmological scenarios. In
section IV, we study analytically the phase space of the model,
locating equilibrium points, discussing their stability and
bifurcations and understanding the flow of the system in the phase
space. Section V contains numerical results from the integration
of the dynamical system, specified for all four mentioned
cosmological scenarios, for which the system is indeed autonomous.
The numerical study qualitatively and quantitatively proves the
results of the previous section, and the viability of the model in
each case is clarified. Finally, in section VI, we provide an
extensive summary of the paper, with a discussion on the results
and their interpretation, along with future prospects of this
work.


\section{The Theoretical Framework of Einstein-Gauss-Bonnet Theory of Gravity}

The canonical scalar field theory of gravity has the following
action,
\begin{equation} \label{eq:Hilbert1}
S = \int d^{4}x \sqrt{-g} \Big( \dfrac{R}{2} - \dfrac{1}{2}
g^{\mu\nu} \nabla_{\mu}\nabla_{\nu} \phi - V(\phi) + L_{matter}
\Big) \, ,
\end{equation}
where $g^{\mu\nu}$ the metric and $\sqrt{-g}$ its determinant, $R
= g^{\mu\nu} R_{\mu\nu}$ the Ricci scalar and $R_{\mu\nu}$ the
Ricci tensor, $\phi = \phi(x^{\alpha})$ the scalar field with
potential $V(\phi)$ and kinetic term $\dfrac{1}{2} g^{\mu\nu}
\nabla_{\mu}\nabla_{\nu} \phi$, and finally $\mathcal{L}_{matter}
= \sqrt{-g} L_{matter}$ the Lagrangian density of the matter
fields.\footnote{The gravitational constant, $\kappa^2
=\frac{1}{8\pi G}= 2.076 \; 10^{-43} s^2 m^{-1} kg^{-1}$ will be
set equal to unity, since we shall use the reduced Planck physical
units $c = \hbar = G = 1$.} Varying this action with respect to
the metric, the Einstein field equations are obtained with
additional source terms, due to the scalar field, and varying them
with respect to the scalar field, and due to the latter's
separability from curvature, an equation of motion for the scalar
field is derived.

Considering the second-order Gauss-Bonnet term, as a means of
accounting for quantum or string corrections to General
Relativity, we can modify the aforementioned action as follows: we
shall consider a minimal coupling of the scalar field to the Ricci
scalar, one that resembles the classical era, and a non-minimal
coupling of the scalar field to the Gauss-Bonnet invariant, one
that resembles the early- or the late-time dynamics of the
Universe. As a consequence, the action shall take the form,
\begin{equation} \label{eq:Hilbert2}
S = \int d^{4}x \sqrt{-g} \Big( \dfrac{R}{2} - \dfrac{1}{2}
g^{\mu\nu} \nabla_{\mu}\nabla_{\nu} \phi - V(\phi) + h(\phi)
\mathcal{G} + L_{matter} \Big) \, ,
\end{equation}
where $\mathcal{G} = R^2 - 4 R_{\mu\nu} R^{\mu\nu} +
R_{\alpha\mu\beta\nu} R^{\alpha\mu\beta\nu}$ is the Gauss-Bonnet
invariant and $h = h(\phi)$ the coupling function.

Varying the action of Eq. (\ref{eq:Hilbert2}) with respect to the
metric, and taking into account the variations of the Ricci
curvature terms,
\begin{equation} \label{eq:Riccivariation}
\begin{split}
\delta \tensor{R}{_{\alpha\beta}} &= \dfrac{1}{2} \Big( 2 \tensor{R}{^{\mu}_{( \alpha |}} \delta \tensor{g}{_{\rho | \beta )}} - 2 \tensor{R}{^{\kappa}_{\alpha}^{\lambda}_{\beta}} \delta \tensor{g}{_{\kappa\lambda}} - \nabla_{\alpha} \nabla_{\beta} \big( \tensor{g}{^{\kappa\lambda}} \delta \tensor{g}{_{\kappa\lambda}} \big) - \nabla^{\mu} \nabla_{\mu} \big( \delta \tensor{g}{_{\alpha\beta}} \big) \Big)  \\
\delta R &= - \tensor{R}{^{\kappa\lambda}} \delta
\tensor{g}{_{\kappa\lambda}} - \nabla^{\mu} \nabla_{\mu} \big(
\tensor{g}{^{\kappa\lambda}} \delta \tensor{g}{_{\kappa\lambda}}
\big) \, ,
\end{split}
\end{equation}
and the corresponding variation of the Gauss-Bonnet term,
\begin{equation} \label{eq:Gaussvariation}
\delta \mathcal{G} = - 2 R \tensor{R}{^{\mu\nu}} \delta
\tensor{g}{_{\mu\nu}} + 8 \tensor{R}{^{\rho\sigma}}
\tensor{R}{^{\mu}_{\rho}^{\nu}_{\sigma}} \delta
\tensor{g}{_{\mu\nu}} + 4 \tensor{R}{^{\mu\nu}}
\nabla^{\kappa}\nabla_{\kappa} \delta \tensor{g}{_{\mu\nu}} - 2
\tensor{R}{^{\kappa\nu\rho\sigma}}\tensor{R}{^{\lambda}_{\nu\rho\sigma}}
\delta \tensor{g}{_{\kappa\lambda}} - 4
\tensor{R}{^{\mu\rho\nu\sigma}} \nabla_{\rho} \nabla_{\sigma}
\delta \tensor{g}{_{\mu\nu}} \, ,
\end{equation}
we get the field equations of the Einstein-Gauss-Bonnet theory,
\begin{equation}\label{eq:Field}
G_{\alpha\beta} = T_{\alpha\beta}^{(matter)} - 2
V(\phi)\tensor{g}{_{\alpha\beta}} +  T_{\alpha\beta}^{(c)} \, ,
\end{equation}
where,
\begin{equation*}
G_{\alpha\beta} = R_{\alpha\beta} - \dfrac{1}{2} R
g_{\alpha\beta}\, ,
\end{equation*}
is the Einstein tensor, and also,
\begin{equation*}
T_{\mu\nu}^{(matter)} = -\dfrac{2}{\sqrt{-g}} \dfrac{\delta
\mathcal{L}_{matter}}{\delta g^{\mu\nu}} = g_{\mu\nu} L_{matter} -
2 \dfrac{\partial L_{matter}}{\partial g^{\mu\nu}}
\end{equation*}
is the energy-momentum tensor associated with the matter fields.
Finally,
\begin{align*}
T_{\alpha\beta}^{(c)} = -& 2 \Big[ h(\phi) \Big( \dfrac{1}{2} \mathcal{G} \tensor{g}{_{\alpha\beta}} + 4\tensor{R}{_{\alpha\mu}}\tensor{R}{_{\beta}^{\mu}} + 4\tensor{R}{^{\mu\nu}}\tensor{R}{_{\alpha\mu\beta\nu}} - 2\tensor{R}{_{\alpha}^{\mu\nu\rho}}\tensor{R}{_{\beta\mu\nu\rho}} - 2 R \tensor{R}{_{\alpha\beta}} \Big) - \\
&- 4\Big(g^{\mu\rho} g^{\nu\sigma} \nabla_{\rho}\nabla_{\sigma} h(\phi) \tensor{R}{_{\alpha\mu\beta\nu}} - g^{\mu\nu} \nabla_{\mu}\nabla_{\nu} h(\phi) \tensor{R}{_{\alpha\beta}} + 2\nabla_{\mu}\nabla_{(\beta} h(\phi) \tensor{R}{^{\mu}_{\alpha)}} - \dfrac{1}{2}\nabla_{\alpha}\nabla_{\beta} h(\phi) R \Big) + \\
&+ 2 \big( \nabla_{\mu}\nabla_{\nu}h(\phi) \tensor{R}{^{\mu\nu}} -
g^{\mu\nu} \nabla_{\mu}\nabla_{\nu} h(\phi) R \big)
\tensor{g}{_{\alpha\beta}} \Big]
\end{align*}
is the pseudo-energy-momentum tensor associated with the
Gauss-Bonnet invariant. Varying with respect to the scalar field,
$\phi$, we obtain the general equation of motion for it,
\begin{equation} \label{eq:Motion}
g^{\alpha\beta}\nabla_{\alpha}\nabla_{\beta}\phi - \dfrac{\partial
V}{\partial \phi} - \dfrac{\partial h}{\partial \phi} \mathcal{G}
= 0 \, .
\end{equation}

The conservation laws completing the picture, are easily obtained,
\begin{align} \label{eq:Conservation}
\nabla_{\alpha} T^{\alpha\beta \; (matter)} &= 0 \;\;\; \text{and} \\
\nabla_{\alpha} T^{\alpha\beta \; (c)} &= 2 g^{\alpha\beta}
\nabla_{\alpha} V(\phi) \, .
\end{align}
The first of these corresponds to the classical law for the
conservation of energy and momentum, the second is an ``energy''
condition imposed by the modification of the spacetime geometry,
namely by the inclusion of the scalar field and the Gauss-Bonnet
invariant.

We consider a flat FRW spacetime with line element,
\begin{equation}\label{eq:FRWmetric}
ds^{2} = -dt^{2} + a(t)^{2} \sum_{i=1,2,3} (dx^{i})^{2} \, ,
\end{equation}
where $a(t)$ is the scale factor, and we assume a torsion-less,
symmetric and metric compatible connection, namely the Levi-Civita
connection,
\begin{equation}\label{eq:FRWconnection}
\tensor{\Gamma}{^{0}_{00}} = 0 \; \, , \;\;\;
\tensor{\Gamma}{^{0}_{i0}} = \tensor{\Gamma}{^{0}_{0i}} = 0 \; \,
, \;\;\; \tensor{\Gamma}{^{0}_{ij}} = \dot{a} a
\tensor{\delta}{_{ij}} \;\;\; \text{and} \;\;\; \text{and} \;\;\;
\tensor{\Gamma}{^{0}_{i0}} = \tensor{\Gamma}{^{0}_{0i}} =
\dfrac{\dot{a}}{a} \tensor{\delta}{_{ij}} \, ,
\end{equation}
where $\tensor{\delta}{_{ij}}$ is the Kronecker tensor. The Ricci
scalar is given as
\begin{equation}\label{eq:FRWricci}
R = 6\dot{H} + 12 H^{2}\, ,
\end{equation}
while the Gauss-Bonnet invariant takes the form,
\begin{equation}\label{eq:FRWgauss}
\mathcal{G} = 24 H^{2} (\dot{H} + H^{2}) \, ,
\end{equation}
where $H = \dfrac{\dot{a}}{a}$ the Hubble expansion rate. We note
here that ``dots'' denote derivatives with respect to the cosmic
time $\Big( \dot{a} = \dfrac{\mathrm{d}a}{\mathrm{dt}} \Big)$,
while ``primes'' will stand for the derivatives with respect to
the e-foldings number, $\Big(a' =
\dfrac{\mathrm{d}a}{\mathrm{dN}}\Big)$.

Furthermore, we consider the contents of the Universe to be
described by ideal fluids, such that the energy-momentum tensor
assumes the simple form
\begin{equation}\label{eq:EnergyMomentum}
T^{00} = \left( \rho_{m} + \rho_{r} \right) \; \, , \;\;\;  T^{0i}
= T^{i0} = 0 \;\;\; \text{and} \;\;\; T^{ij} = -\left( P_{m} +
P_{r} \right) \delta^{ij} \, ,
\end{equation}
where $\rho_{m}$ is the mass-energy density and $P_{m}$ the
pressure of non-relativistic fluids (matter) , while $\rho_{r}$ is
the mass-energy density and $P_{r}$ the pressure of relativistic
fluids (radiation). According to current theoretical approaches,
the Universe was dominated by these two kinds of fluids and these
two shall be used in our analysis as well. The equation of state
concerning matter is that of pressureless dust
\begin{equation}
P_{m} = 0 \, ,
\end{equation}
while the one corresponding to radiation is
\begin{equation}
P_{r} = \dfrac{1}{3} \rho_{r} \, .
\end{equation}
A short comment is necessary here, since the Universe is easily
said to contain many different matter fields. Speaking of
radiation, we refer to any kind of matter that reaches very high
energies (\textit{e.g.} light, neutrinos, electrons, positrons,
\textit{etc.}). In the same manner, dust stands for any kind of
matter that has very low energy (such as typical baryonic matter).
From this perspective, hot dark matter can be considered a form of
radiation, while cold dark matter is generally included in dust.

The field equations of the theory for the FRW metric, that is Eqs.
(\ref{eq:Field}), reduce to the Friedmann equation,
\begin{equation} \label{eq:FRWfield1}
3 H^{2} - \dfrac{1}{2} \dot{\phi}^2 - V(\phi) - 24 \dfrac{\partial
h}{\partial \phi} \dot{\phi} H^{3} = \rho_{m} + \rho_{r} \, ,
\end{equation}
and the Landau-Raychaudhuri equation,
\begin{equation} \label{eq:FRWfield2}
\dot{H} - \dot{\phi}^2 - 8 \Big( H^{2} \dot{\chi}
\dfrac{\partial^{2} h}{\partial \chi^{2}} + H^{2} \ddot{\phi}
\dfrac{\partial h}{\partial \phi} + 2 H (\dot{H}+H^{2}) \dot{\phi}
\dfrac{\partial h}{\partial \phi} \Big) = \rho_{m} + P_{m} +
\rho_{r} + P_{r} \, .
\end{equation}
In the same manner, the equation of motion for the scalar field,
that is, Eq. (\ref{eq:Motion}), becomes,
\begin{equation} \label{eq:FRWmotion}
\ddot{\phi} + 3 H \dot{\phi} + 3\dot{\phi} \dfrac{\partial
V}{\partial \phi} + 24 \dfrac{\partial h}{\partial \phi} H^{2}
(\dot{H} + H^{2} ) = 0 \, .
\end{equation}
Finally, the conservation laws for energy and momentum, namely
Eqs. (\ref{eq:Conservation}), are reduced to the continuity
equations for the mass-energy densities of matter and radiation,
which are,
\begin{equation}\label{eq:Continuity}
\begin{split}
\dot{\rho}_{m} + 3H \left( \rho_{m} + 3 P_{m} \right) &= \dot{\rho}_{m} + 3H \rho_{m} = 0 \;\;\; \text{and} \\
\dot{\rho}_{r} + 3H \left( \rho_{r} + 3 P_{r} \right) &=
\dot{\rho}_{r} + 4H \rho_{r} = 0 \, .
\end{split}
\end{equation}
Another specification we need to make concerns the form of the
potential and that of the coupling function. Several cases in the
literature (see Refs. \cite{Bamba:2007ef, Nojiri:2019dwl} for
example) convince us that the exponential form is a good
approximation for both of them. As a result, we may consider the
potential to be a decreasing function of the scalar field in the
form
\begin{equation} \label{eq:Potential}
V(\phi) = V_{0} e^{-\lambda\phi} \, ,
\end{equation}
where $V_{0} > 0$ and $\lambda > 0$ two positive constants, so
that
\begin{equation*}
\dfrac{\partial V}{\partial\phi} = -\lambda V_{0} e^{-\lambda\phi}
= -\lambda V(\phi) < 0 \, ,
\end{equation*}
and the coupling function to be increasing in the form
\begin{equation} \label{eq:Coupling}
h(\phi) = \dfrac{h_{0}}{\mu} e^{\mu\phi} \, ,
\end{equation}
where $h_{0} > 0$ and $\mu > 0$ two positive constants, so that
\begin{equation*}
\dfrac{\partial h}{\partial\phi} = h_{0} e^{\mu\phi} = \mu h(\phi)
> 0 \;\;\; \text{and} \;\;\; \dfrac{\partial^2 h}{\partial\phi^2}
= \mu h_{0} e^{\mu\phi} = \mu^2 h(\phi) > 0 \, .
\end{equation*}
Thus, the larger the scalar field is, the weaker its potential
becomes and the stronger its coupling to the Gauss-Bonnet
invariant. This essentially means that, in regimes where the
potential has great effect in the scalar field (the field acts
dynamically), the quadratic curvature terms are small and possibly
negligible, while in regimes where the scalar field does not
evolve dynamically, the quadratic curvature terms dominate.

\section{The Autonomous Dynamical System of Einstein-Gauss-Bonnet Theory}

In order to examine the cosmological implications of these models
by means of their phase space, we need to define dimensionless
phase space variables. Considering that totally five elements are
active in the theory, namely the kinetic term of the scalar field,
the potential of the scalar field, the Gauss-Bonnet term and the
two components of the cosmic fluid (matter and radiation), we may
define the following five dynamical variables
\begin{equation} \label{eq:PhaseVar}
x_{1} = \dfrac{1}{H} \sqrt{\dfrac{\dot{\phi}^2}{6}} \;\; \, ,
\;\;\; x_{2} = \dfrac{1}{H} \sqrt{\dfrac{V(\phi)}{6}} \;\; \, ,
\;\;\; x_{3} = H^2 \dfrac{\partial h}{\partial \phi} \;\; \, ,
\;\;\; x_{4} = \dfrac{1}{H} \sqrt{\dfrac{\rho_{r}}{3}} \;\;\;
\text{and} \;\;\; x_{5} = \dfrac{1}{H} \sqrt{\dfrac{\rho_{m}}{3}}
\, .
\end{equation}
As said, the first two of these variables involve the scalar
field, the third involves the quadratic gravity (or rather its
coupling to the scalar field), and the last two involve the matter
fields.

Their evolution will be studied with respect to the $e$-foldings
number, $N$, defined as
\begin{equation} \label{eq:e-folds}
N = \int_{t_{in}}^{t_{fin}} H(t) dt \, ,
\end{equation}
where $t_{in}$ and $t_{fin}$ the initial and final moments of the
coordinate time. This transpose from the coordinate time to the
$e$-foldings number is a necessary transformation, since eras of
rapid evolution -such as the early- or late-time accelerated
expansions- are easier to study when physical quantities are
weighted with the Hubble rate. The derivatives with respect to the
$e$-foldings number are derived from the derivatives with respect
to the coordinate time, as,
\begin{equation*}
\dfrac{\mathrm{d}}{\mathrm{dN}} = \dfrac{1}{H}
\dfrac{\mathrm{d}}{\mathrm{dt}} \;\; \text{and} \;\;\;
\dfrac{\mathrm{d}^2}{\mathrm{dN}^2} = \dfrac{1}{H} \Big(
\dfrac{\mathrm{d}^2}{\mathrm{dt}^2} - \dfrac{\dot{H}}{H}
\dfrac{\mathrm{d}}{\mathrm{dt}} \Big) \, .
\end{equation*}
In order to derive the equations of evolution for the dynamical
variables (\ref{eq:PhaseVar}), we shall use the equations of
motion of motion for the five respective elements of the
cosmological models, more specifically the Friedmann and
Raychaudhuri equations (Eqs. (\ref{eq:FRWfield1}) and
(\ref{eq:FRWfield2})), the Klein-Gordon equation (Eq.
(\ref{eq:FRWmotion})) and the two continuity equations (Eqs.
(\ref{eq:Continuity})). Specifically, the evolution of $x_{1}$
with respect to the $e$-foldings number is given as
\begin{equation}
\dfrac{\mathrm{d} x_{1}}{\mathrm{dN}} = \dfrac{1}{H} \dot{x}_{1} =
\dfrac{1}{H} \sqrt{\dfrac{\dot{\phi}^2}{6}} \Bigg(
\dfrac{\ddot{\phi}}{H \dot{\phi}} - \dfrac{\dot{H}}{H^2} \Bigg) \,
,
\end{equation}
where, from the field equations (Eq. (\ref{eq:FRWmotion})),
\begin{equation*}
\dfrac{\ddot{\phi}}{H \dot{\phi}} = -3 - \dfrac{1}{H}
\dfrac{\partial V}{\partial \phi} - \dfrac{\partial h}{\partial
\phi} H \left( \dot{H} - H^2 \right) \, ,
\end{equation*}
and from the definition of the dynamical variables,
\begin{equation*}
-\dfrac{1}{H} \dfrac{\partial V}{\partial \phi} =
\dfrac{\sqrt{6}}{2} \lambda \dfrac{x_{2}^2}{x_{1}}  \;\;\;
\text{and} \;\;\; \dfrac{\partial h}{\partial \phi} H \left(
\dot{H} - H^2 \right) = 4\sqrt{6} \dfrac{x_{3}}{x_{1}} \Big( 1 - m
\Big) \, ,
\end{equation*}
where $m = \dfrac{\dot{H}}{H^2}$. Consequently, the first
differential equation is
\begin{equation} \label{eq:firstODE}
x_{1}^{\prime} = - 3x_{1} + \dfrac{\sqrt{6}}{2}\lambda x_{2}^2 -
4\sqrt{6} x_{3} - \big( x_{1} + 4\sqrt{6} x_{3} \big)m \, .
\end{equation}

The evolution of $x_{2}$ with respect to the $e$-foldings number
is given as
\begin{equation}
\dfrac{\mathrm{d} x_{2}}{\mathrm{dN}} = \dfrac{1}{H} \dot{x}_{2} =
\dfrac{1}{H} \sqrt{\dfrac{V(\phi)}{3}} \Big( \dfrac{1}{H}
\dfrac{\dot{V}(\phi)}{V(\phi)} - \dfrac{\dot{H}}{H^2} \Big) \, ,
\end{equation}
where, from the chain rule of differentiation,
\begin{equation*}
\dot{V}(\phi) = \dfrac{\partial V}{\partial \phi} \dot{\phi} \, ,
\end{equation*}
and from the definitions of the dynamical variables,
\begin{equation*}
\dfrac{1}{H} \dfrac{\dot{V}(\phi)}{V(\phi)} =
\dfrac{1}{H}\dfrac{\partial V}{\partial \phi}
\dfrac{\dot{\phi}}{V(\phi)} = -\dfrac{\sqrt{6}}{2} \lambda x_{2}
\, .
\end{equation*}
Consequently, the second differential equation is,
\begin{equation} \label{eq:secondODE}
x_{2}^{\prime} = -x_{2} \Big( \dfrac{\sqrt{6}}{2} \lambda x_{1} +
m \Big) \, .
\end{equation}
Following, the evolution of $x_{3}$ with respect to the
$e$-foldings number is given as,
\begin{equation}
\dfrac{\mathrm{d} x_{3}}{\mathrm{dN}} = \dfrac{1}{H} \dot{x}_{3} =
H \Big( \dfrac{\partial h}{\partial \phi} \Big)^{\cdot} + 2\dot{H}
\dfrac{\partial h}{\partial \phi} \, .
\end{equation}
Again using the chain  rule of differentiation, we obtain,
\begin{equation*}
\Big( \dfrac{\partial h}{\partial \phi} \Big)^{\cdot} =
\dfrac{\partial^2 h}{\partial \phi^2} \dot{\phi} = \mu
\dfrac{\partial h}{\partial \phi} \dot{\phi} \, ,
\end{equation*}
and from the definitions of the dynamical variables we get,
\begin{equation*}
\mu H \dfrac{\partial h}{\partial \phi} \dot{\phi} =
\dfrac{\sqrt{6}}{2}\mu x_{1} x_{3}\, .
\end{equation*}
Consequently, the third differential equation is,
\begin{equation} \label{eq:thirdODE}
x_{3}^{\prime} = x_{3} \Big( \dfrac{\sqrt{6}}{2}\mu x_{1} + m
\Big) \, .
\end{equation}
The evolution of $x_{4}$ and that of $x_{5}$ with respect to the
$e$-foldings number are identical, just derived from slightly
different continuity equations; they are given as,
\begin{align}
\dfrac{\mathrm{d} x_{4}}{\mathrm{dN}} &= \dfrac{1}{H} \dot{x_{4}} = \dfrac{1}{H} \sqrt{\dfrac{\rho_{r}}{3}} \Bigg( \dfrac{\dot{\rho}_{r}}{2H \rho_{r}} - \dfrac{\dot{H}}{H^2} \Bigg) \;\; \text{and} \\
\dfrac{\mathrm{d} x_{5}}{\mathrm{dN}} &= \dfrac{1}{H} \dot{x_{5}}
= \dfrac{1}{H} \sqrt{\dfrac{\rho_{m}}{3}} \Bigg(
\dfrac{\dot{\rho}_{m}}{2H \rho_{m}} - \dfrac{\dot{H}}{H^2} \Bigg)
\, .
\end{align}
Given the continuity equations (\ref{eq:Continuity}), we may
write,
\begin{equation*}
\dfrac{\dot{\rho_{r}}}{2H \rho_{r}} = -2 \;\;\; \text{and} \;\;\;
\dfrac{\dot{\rho_{m}}}{2H \rho_{m}} = -\dfrac{3}{2} \, ,
\end{equation*}
and employing the definitions of the dynamical variables, the
fourth differential equations becomes,
\begin{equation} \label{eq:fourthODE}
x_{4}^{\prime} = -x_{4} \big( 2 + m \big) \, ,
\end{equation}
while the fifth differential equations takes the form,
\begin{equation} \label{eq:fifthODE}
x_{5}^{\prime} = -x_{5} \Big( \dfrac{3}{2} + m \Big) \, .
\end{equation}
We should notice that the last two differential equations are
separable from the first three and analytically integrated. As a
result, the actual matter fields do not interact with the scalar
field or the Gauss-Bonnet invariant and evolve independently from
the latter. On the other hand, due to their coupling, the scalar
field and the Gauss-Bonnet invariant cannot evolve independently,
as one can easily understand this by merely looking at the
intermingling of the first three differential equations. We should
notice that radiation and matter could interact with each other,
which would cause significant change in Eqs. (\ref{eq:fourthODE})
and (\ref{eq:fifthODE}) -probably breaking their integrability.
This interaction, however, would not affect the remaining three
variables, as Eqs. (\ref{eq:firstODE}-\ref{eq:thirdODE}) would not
be altered. Notably, the complete phase space of the theory is
separated to two linearly independent subspaces, one concerning
the evolution of the scalar field and the quadratic curvature
terms, and one concerning the evolution of the cosmic fluids. It
has already become clear that the latter is trivial and rather
indifferent, whereas the former contains all the necessary
information about the development of the cosmological models.
Consequently, we shall deal with the reduced three-dimensional
phase space of $x_{1}$, $x_{2}$ and $x_{3}$; it is, of course,
necessary to point out that this reduced phase space, although it
contains the fundamental structure of the Einstein-Gauss-Bonnet
theory, does not account for the complete evolution of the
Universe, since the matter fields are absent.

Let us quote here the final form of the dynamical system for
clarity,
\begin{align}\label{claritydynsystem}
& x_{1}^{\prime} = - 3x_{1} + \dfrac{\sqrt{6}}{2}\lambda x_{2}^2 -
4\sqrt{6} x_{3} - \big( x_{1} + 4\sqrt{6} x_{3} \big)m \, , \\
\notag & x_{2}^{\prime} = -x_{2} \Big( \dfrac{\sqrt{6}}{2} \lambda
x_{1} + m \Big) \, , \\ \notag & x_{3}^{\prime} = x_{3} \Big(
\dfrac{\sqrt{6}}{2}\mu x_{1} + m \Big) \, , \\ \notag &
x_{4}^{\prime} = -x_{4} \big( 2 + m \big) \, , \\ \notag &
x_{5}^{\prime} = -x_{5} \Big( \dfrac{3}{2} + m \Big) \, .
\end{align}
Let us discuss in brief our strategy for studying the phase space.
Basically, our aim is to study in detail solution subspaces of the
whole phase space corresponding to the dynamical system
(\ref{claritydynsystem}), with certain cosmological interest. The
study of the whole phase space would be quite difficult to study
analytically, because the parameter $m$ would be cosmic
time-dependent, thus the dynamical system would be non-autonomous.
However, if the parameter $m$ is constant, then the dynamical
system is rendered autonomous, and it can be safely studied for
several different values of the parameter $m$. Due to the fact
that $m$ may take different values, this will certainly affect the
stability properties of the dynamical system. More importantly, we
should not that the parameters $\mu$ and $\lambda$ are free
parameters of the theory, and by no means we do not fix their
value. The values of $\mu$ and $\lambda$ will affect the stability
of the subspaces of the phase space corresponding to different
values of the parameter $m$. For example, for a specific value of
$m$, the values of $\mu$ and $\lambda$ may make the fixed points
of the phase space corresponding to $m$, stable or unstable. In
some sense, the values of $\mu$ and $\lambda$ may indicate
attractors in the phase space, for the various values of $m$.

It is worth to show explicitly the cases of interest which we
shall study in the following sections, which correspond to
$m=$constant. Recall that the parameter $m$ is defined to be
$m=\frac{\dot{H}}{H^2}$, so if $m$ is constant, the differential
equation yields the solution,
\begin{equation}\label{powerlawsolutionhubble}
H(t)=\frac{1}{\Lambda-m t}\, ,
\end{equation}
where $\Lambda$ is an integration constant. The Hubble rate of Eq.
(\ref{powerlawsolutionhubble}) corresponds to the scale factor,
\begin{equation}\label{scalefactor}
a(t)=\beta (\Lambda -m t)^{-1/m}\, ,
\end{equation}
where $\beta$ is an integration constant. Thus it is a power-law
type scale factor, and it is apparent that our study is devoted in
studying a subspace of solutions of the whole phase space, which
corresponds to solutions with power-law types of scale factors.
For example, the case $m=0$ corresponds to a de Sitter solution
with $H=\frac{1}{\Lambda}$. If $m = -3/2$ and $\Lambda=0$, we have
a matter dominated evolution with
$a(t)=\left(\frac{3}{2}\right)^{2/3} \beta  t^{2/3}$, while for $m
= -2$ we have a radiation dominated evolution $a(t)=\sqrt{2} \beta
t^{1/2}$. Finally, for $m = -3$ we have a stiff evolution with
$a(t)= \beta 3^{1/3}t^{1/3}$.

Finally, we need to note that if for some of the above constant
values of $m$, if some values of the free parameters  $\mu$ and
$\lambda$ make the fixed points of the dynamical system stable,
the corresponding solutions are attractors of the phase space, and
thus could be viewed as some sort of tracking solutions. Also, the
examination of the phase space for general $m$ would simply be
impossible to study analytically, because the study of a
non-autonomous dynamical system is not simple at all, one must
find limiting cycles in order to claim, without certainty, that
attractors exist. Thus the only way we found that yields
analytical results, is to study subspaces of the phase space that
yield constant parameter $m$, and also correspond to several
power-law types of scale factor. Happily, most of the cosmological
eras of interest are cases with constant $m$, and this is what we
exploited in this paper in order to perform the study.

\subsection{Integrability of two Ordinary Differential Equations}

The dynamical system of the Einstein-Gauss-Bonnet theory can only
be autonomous for the case the parameter $m$ is a constant. This
constraint can severely reduce the number of cosmological
scenarios, since it can be constant for specific cases only.
Remembering that the Universe is composed of barotropic fluids, we
may consider that $P_{eff} = w \rho_{eff}$, where $w$ is the
effective barotropic index of the Universe; we may furthermore
consider, as an \textit{ansatz}, that this index remains constant
for some specific cosmological eras, which are proved to be of
great importance. Such cases identify with de Sitter expansion ($w
= -1$, hence $m = 0$), matter dominated era ($w = 0$, hence $m =
-3/2$), radiation dominated era ($w = 1/3$ , hence $m = -2$) and
stiff-matter era ($w = 1$, hence $m = -3$).\footnote{Given the
standard solutions to FRW model, the Hubble rate is $H(t) =
\dfrac{2}{3(1+w)t}$, hence $m = -\dfrac{3(1+w)}{2}$; if the
barotropic index is constant for specific eras, then $m$ is also
constant.} Note that this is the only consistent way for the
dynamical system of the Einstein-Gauss-Bonnet theory to be
autonomous, a prerequisite for the study of the phase space that
leads to the existence of fixed points or other invariant
structures.

It is not difficult to solve the differential equations
(\ref{eq:fourthODE}) and (\ref{eq:fifthODE}), and their solutions
are given in closed form as exponential functions of the form,
\begin{align}\label{eq:x4x5Sol}
x_{4}(N) &= x_{4(0)} e^{-( 2+m ) (N-N_{0})} \;\; \text{and} \\
x_{5}(N) &= x_{5(0)} e^{-\left( \frac{3}{2} + m \right) (N-N_{0})}
\, ,
\end{align}
proposing that $m = \dfrac{\dot{H}}{H^2}$ is a constant and
defining $x_{4(0)}$ and $x_{5(0)}$ to be values of $x_{4}$ and
$x_{5}$ at some point of time $N_{0}$, usually perceived as the
present.

Suppose that $\dfrac{3}{2} > -m$, this would mean that $h(\phi)$
and $V(\phi)$ which are exponentials, are decreasing over the
$e$-foldings number. In the same manner, both exponentials are
increasing over the $e$-foldings number if $2 < -m$. In the middle
interval, $\dfrac{3}{2} < -m < 2$, $x_{4}$ is increasing, while
$x_{5}$ is decreasing. These behaviors are typical in the first
case for the mass-energy density of both matter and radiation,
since they are known to decrease over the scale factor, however,
they seem fully unphysical in the second and the third. Yet,
$x_{4}$ and $x_{5}$ are not the mass-energy densities of the two
fluids, but rather the (square roots of the) corresponding density
parameters. Furthermore, it is a result given with respect to the
$e$-foldings number and may differ with respect to time, depending
on the behavior of the Hubble rate. As a result, we should
transpose both of them as functions of time and then transform
them to the respective mass-energy densities.

We assume a Hubble rate in the form of $H(t) =
\dfrac{2}{3(1+w)t}$, as in typical solutions of the FRW models
(where $w = \frac{1}{3}$ corresponds to radiation and $w = 0$ to
matter). The $e$-foldings number with respect to time is given as
\begin{equation*}
N - N_{0} = \dfrac{2}{3(1+w)} \ln \big( 3(1+w)t \big) \, ,
\end{equation*}
and we may define $t_{0}$ as the instance of coordinate time that
corresponds to $N_{0}$ -the present time. Substituting this to
Eqs. (\ref{eq:x4x5Sol}), we obtain,
\begin{align*}
x_{4} &= x_{4(0)} \left(3(1+w)t\right)^{-\frac{2(2+m)}{3(1+w)}}  \;\; \text{and} \\
x_{5} &= x_{5(0)} \left(3(1+w)t\right)^{-\frac{3+2m}{3(1+w)}} \, ,
\end{align*}
and transforming them to mass-energy densities, by means of their
definitions from Eqs. (\ref{eq:PhaseVar}), we can see that,
\begin{align*}
\rho_{r} = \rho_{r(0)} \Bigg(\dfrac{t_0}{t}\Bigg)^2 \left(3(1+w)t\right)^{-\frac{4(2+m)}{3(1+w)}} \;\; \text{and} \\
\rho_{m} = \rho_{m(0)} \Bigg(\dfrac{t_0}{t}\Bigg)^2
\left(3(1+w)t\right)^{-\frac{6+4m}{3(1+w)}} \, ,
\end{align*}
where $\rho_{r(0)}$ and $\rho_{m(0)}$ the mass-energy densities at
the moment $t_{0}$. Assuming $w = \dfrac{1}{3}$, so $m = -2$, for
$x_{4}$, and $w=0$, so $m = -\dfrac{3}{2}$, for $x_{5}$, we see
that the exponential vanishes and the remaining terms signify the
familiar $\sim t^{-2}$ behavior of the mass-energy densities in
the FRW models.

What is interesting here, is the fact that both dynamical
variables are decreasing from some initial values towards their
equilibrium point $x_{4}^{*} = x_{5}^{*} = 0$. This convergence
towards the equilibrium point may be slow, depending on the
magnitude of $m$, but it is relatively fast and absolutely
certain. Combined with the independent evolution of $x_{4}$ and
$x_{5}$ from the other three variables, we could discard the
corresponding differential equations and restrict our focus on the
afore-mentioned reduced phase space of the model, that contains
only $x_{1}$, $x_{2}$ and $x_{3}$, namely the geometric features
of the Einstein-Gauss-Bonnet theory. The evolution of these three
variables takes place in the other subspace and is not at all
affected by the evolution of $x_{4}$ and $x_{5}$. At first glance,
this simplification bears the physical meaning of a Universe empty
of matter fields, yet this is not at all the case. What happens is
that the system will generally stabilize on $x_{4}^{*} = x_{5}^{*}
= 0$, hence the behavior of the matter fields does not affect us
during the evolution of the remaining three variables; this allows
us to study how $x_{1}$, $x_{2}$ and $x_{3}$ evolve and whether
they settle to a fixed point, independently of $x_{4}$ and
$x_{5}$. Furthermore, the result will be easier visualized and
concentrated on the distinct elements of the theory -that is the
scalar field and the quadratic curvature.

As we noted, in order for all these to be consistent, the
parameter $m$ must be a constant and take specific values of
physical interest. These values are justified, as we will discuss
further in section III.C. However, in order to further clarify
things in a correct way, let us briefly discuss here the choice of
the dynamical system variables and also the features of the
parameter $m$. The purpose of the dynamical system is indeed to
study the cosmological aspects of the respective theory, in
accordance to many similar approaches. The choice of the dynamical
variables is made as such to incorporate the following:
\begin{itemize}
    \item the proper rescaling, according to the Friedmann equation,
    \item the shorting of the most interesting variables of the
Einstein-Gauss-Bonnet theory, and
    \item the discrimination between different phases of
cosmological evolution.
\end{itemize}
As a result, we consider the theory to contain the following five
(and not four) degrees of freedom:
\begin{itemize}
\item[1] The kinetic term of the scalar field,
$\dfrac{\dot{\phi}}{2}$, that also stands for the evolution of the
scalar field $\phi$ itself; this is reflected on $x_{1}$ \item[2]
The potential of the scalar field, $V(\phi)$, that causes the
dynamical evolution of the $\phi$; it is reflected on $x_{2}$.
\item[3] The emergence of the Gauss-Bonnet invariant and its
coupling to the scalar field, as measured by the coupling function
$h(\phi)$; this is reflected on $x_{3}$. \item[4] The presence of
matter fields, namely non-relativistic matter (radiation) and
relativistic matter (dust); these are reflected on $x_{4}$ and
$x_{5}$.
\end{itemize}
The Hubble rate, $H$, is used for the proper rescaling of the
afore-mentioned variables and of the $e$-foldings number and not
as a variable itself, since it cannot considered as a separate
degree of freedom. In fact, adding the Hubble rate as one (say
$x_{6} = \dfrac{\dot{H}}{H^2}$) would probably cause an
over-determination problem; the evolution of this variable is
given through the Raychaudhuri equation, also used for the
evolution of $x_{3}$, so a certain degeneracy would emerge in the
dynamical system. It is worth mentioning that this
re-parametrization has been followed by other authors as well,
most notably in the book by \cite{tsuji}.

The direct way of dealing with the cosmological system is to
actually solve for $\dfrac{\dot{H}}{H^2}$, however, this is a
rather unmanageable situation. The Hubble rate and its derivative
over time are linked to the Ricci scalar, hence they are
originally acquired from the field equations. These are a system
of ten non-linear partial differential equations, whose analytical
solution is not generally obtained. What we did was to assume a
specific form of the solution (the Robertson-Walker metric) and
reduce them to a system of five non-linear ordinary differential
equations, much more easy to treat qualitatively and
quantitatively; this system is the one re-parametrized by Eqs.
(25), (27), (29), (32) and (33). Attempting to solve the original
system, we could account for an endogenous evolution of the Hubble
rate, yet this is not the task we set for ourselves.

Having reduced the field equations to an extended
Friedmann-Robertson-Walker system, we assume that the Hubble rate
will evolve according to the standard solutions of the classical
theory, namely $H = H_0$ for the de Sitter expansion and $H =
\dfrac{2}{3(1+w)t}$ for the case of some barotropic fluid, whereas
$\dfrac{P}{\rho} = w$. In both cases, $\dfrac{\dot{H}}{H^2}$ is a
constant, proposing the barotropic index is constant; the latter
is known to happen for the distinct case we are dealing with,
cases that are identified as specific eras in the evolution of the
Universe.

Consequently, our choice to assume $m = \dfrac{\dot{H}}{H^2}$ to
be constant, one that we will treat as exogenously set, is an
\textit{ansatz} based on the previous remarks. Indeed this
\textit{ansatz} reduces the generality of the model, but accounts
for specific stages in the evolution of the Universe, in
accordance to the Standard Cosmological Model. Furthermore, it
allows the system of Eqs. (25), (27), (29), (32) and (33) to be
autonomous, without any additional equations -that would cause
degeneracy- or assumptions -that would further reduce the system.
Of course, if some of the constant $m$ solutions, are not actual
solutions of the cosmological system, these would not lead to
fixed points of the dynamical system. As we show later on, both
these solutions lead to fixed points that generally satisfy the
Friedman constraint.


\subsection{The Friedmann constraint}

Considering all ingredients of the universe as homogeneous ideal
fluids, we may rewrite Eq. (\ref{eq:FRWfield1}) as,
\begin{equation*}
\dfrac{\dot{\phi}^2}{6 H^2} - \dfrac{V(\phi)}{3 H^2} + 8
\dfrac{\dot{\phi}}{H^2} \dfrac{\partial h}{\partial \phi} +
\dfrac{\rho_{r}}{3 H^2} + \dfrac{\rho_{m}}{3 H^2} = 1 \, ;
\end{equation*}
defining the density parameters for each component of the cosmic
fluid, either actual or effective,
\begin{equation*}
\Omega_{\phi} = \dfrac{\dot{\phi}^2}{6 H^2} - \dfrac{V(\phi)}{3
H^2} \; \, , \;\;\; \Omega_{GB} = 8 \dfrac{\dot{\phi}}{H^2}
\dfrac{\partial h}{\partial \phi} \; \, , \;\;\; \Omega_{r} =
\dfrac{\rho_{r}}{3 H^2} \;\;\; \text{and} \;\;\; \Omega_{m} =
\dfrac{\rho_{m}}{3 H^2}\, .
\end{equation*}
In effect we have a dimensionless form of the Friedmann equation,
\begin{equation}
\Omega_{\phi} + \Omega_{GB} + \Omega_{r} + \Omega_{m} = 1 \, .
\end{equation}
Taking into consideration the definitions of the phase space
variables, from Eqs. (\ref{eq:PhaseVar}), we can see that,
\begin{equation*}
\Omega_{\phi} = x_{1}^2 + x_{2}^2 \; \, , \;\;\; \Omega_{GB} =
8\sqrt{6} x_{1} x_{3} \; \, , \;\;\; \Omega_{r} = x_{4}^2 \;\;\;
\text{and} \;\;\; \Omega_{m} = x_{5}^2 \, .
\end{equation*}
Substituting in the dimensionless form of the Friedmann equation,
we eventually obtain the Friedmann constraint,
\begin{equation} \label{eq:FriedConstraint}
x_{1}^2 + x_{2}^2 + 8\sqrt{6} x_{1} x_{3} + x_{4}^2 + x_{5}^2 = 1
\, .
\end{equation}
This constraint naturally restricts the five dynamical variables
on a hypersurface of the phase space, so they can take specific
values in correspondence to each other. This essentially means
that the dimension of the phase space is reduced from five to
four.

\begin{figure}
\includegraphics[width=0.5\linewidth]{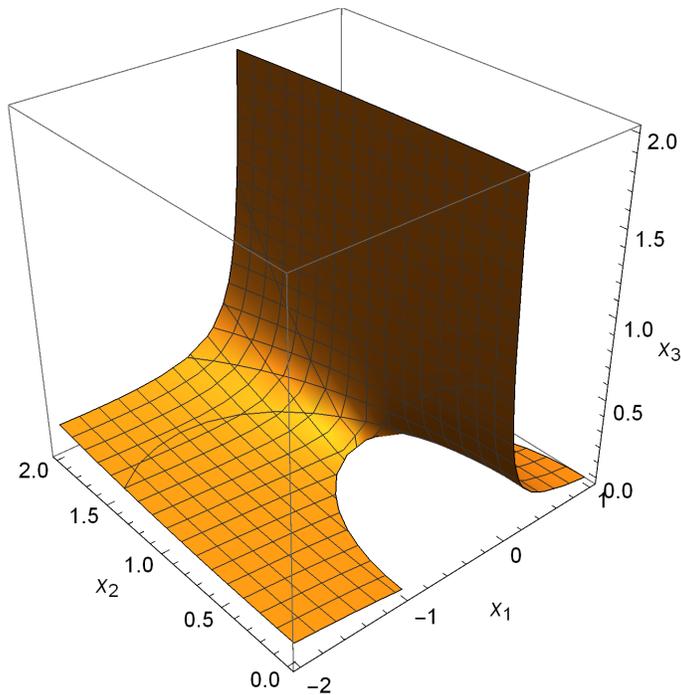}
\caption{The $2-d$ surface of the Friedmann constraint on the
phase space of $x_1$, $x_2$ and $x_3$.}\label{fig:FriedConstraint}
\end{figure}

The physical meaning of the constraint is derived from the
flatness of the Universe. If the Universe is indeed described by a
flat FRW metric -as observations indicate- and contains nothing
else than what we already considered, then the fulfillment of the
Friedmann constraint is a prerequisite. Its non-fulfillment, on
the other hand, would hint as to a non-flat $3-d$ space, or to
missing components in the theory, or from a physical point of
view, non viability of the corresponding solutions. Setting the
latter aside, as we are interested in this specific theory, we
should notice that $x_{1}^2 + x_{2}^2 + 8\sqrt{6} x_{1} x_{3}
x_{4}^2 + x_{5}^2 < 1$ means a FRW Universe with negative spatial
curvature (closed FRW cosmologies), whereas $x_{1}^2 + x_{2}^2 +
8\sqrt{6} x_{1} x_{3} x_{4}^2 + x_{5}^2 < 1$ means one with
negative spatial curvature (open FRW cosmologies).

Generally, we would expect solutions of the system
(\ref{eq:firstODE}-\ref{eq:fifthODE}) to fulfill the Friedmann
constraint at all times, otherwise we would regard them as
non-physical and unviable. However, we shall see that in the
reduced three-dimensional phase space (where $x_{4} = x_{5} = 0$)
not all solutions respect this constraint. In fact, even when a
stable equilibrium point fulfills the constraint, not all
trajectories leading to it, fulfill it as well. This could mean
that the Einstein-Gauss-Bonnet Universe does not fulfill the
Friedmann constraint at all times, and thus it is not described by
flat FRW metric throughout all its history, but only at the final
attractors. It could also mean, on the other hand, that the
assumption of a Universe empty of matter fields is misleading, as
we pointed out, since it could guarantee the fulfillment of the
Friedmann constraint all the way from the ``unviable'' initial
conditions to the viable fixed point. We discuss and investigate
in detail these issues in the following sections.

\subsection{The Free Parameters of the Model}

The model presented so far contains three free parameters. Two of
them are inherited from the assumption of an exponential potential
($\lambda$) and an exponential coupling function ($\mu$),
therefore their magnitude and sign follow the following
restrictions, namely, they are defined as positive and we know
that they must be small and of similar size.\footnote{We will see
afterwards that further restrictions may exist.}

Apart from these, one more free parameter of crucial importance,
is the following,
\begin{equation*}
m = \dfrac{\dot{H}}{H^2} \, .
\end{equation*}
We should remark that this ratio is not generally constant, since
the Hubble rate cannot be known in a closed form \textit{a
priori}, that is before the field equations of the theory are
solved for specific sources. However, we may notice that a great
number of FRW cosmologies -used in both relativistic and modified
theories- yield such forms of the Hubble rate, so that this ratio
is indeed constant in specific regimes, in other words during
specific cosmological eras. Essentially, a constant Hubble rate
(as in the de Sitter case) or one such that is $\sim t^{-1}$
(matter- or radiation-dominated eras) will yield a constant $m$,
if the barotropic fluids of the Universe are specified, as proved
in Section III.A. This can justify our \textit{ansatz} that this
ratio is treated as constant and a free parameter of the system.
Our intention is to study the system for different values of $m$,
corresponding to different overall behaviors of the space
expansion (different cosmological eras), while transitions from
one behavior to the next could be perceived as bifurcations in the
system.

Another point we need to stress, is the magnitude of the parameter
$m$, since the matter and radiation density parameters must be
decreasing functions of the $e$-foldings number, as we stated
earlier. This results to the restriction $m > -\dfrac{3}{2}$. In
fact, it is easy to show that the actual restriction for the
values of $m$ is the interval $[ -3 , 0 ]$. This arises naturally,
if we consider the four fundamental solutions of the Friedmann
equations in the classical case:
\begin{enumerate}
\item Given $w = -1$ or $H(t) = H_0$, that corresponds to a de
Sitter expanding Universe, then it is easy to calculate that $m =
0$. \item Given $w = \dfrac{1}{3}$ or $H(t) = \dfrac{1}{2t}$, that
corresponds to a Universe containing relativistic fluids, or the
radiation-dominated era, then $m = -2$.

\item Given $w = 0$ or $H(t) = \dfrac{2}{3t}$, that corresponds to
a Universe containing non-relativistic (dust) fluids, or the
matter-dominated era, then $m = -\dfrac{3}{2}$.

\item Given $w = 1$ or $H(t) = \dfrac{1}{3t}$, corresponding to a
Universe containing stiff fluids, another era that could be
encountered prior or after the inflation, then $m = -3$.

\end{enumerate}
Containing our analysis to these four cases, not only can we
discuss all (possible) stages of the Universe evolution in the
Standard Cosmological model, but we can also rest assured that $m$
is indeed a constant.

We need to note that various values of the free parameters $\mu$
and $\lambda$ may yield stability or instability of the fixed
points corresponding to the various values of the parameter $m$.
This fact however may seems to indicate that the values of $\mu$
and $\lambda$ may vary during the various cosmological eras. We
need to clarify this issue however, since the phase space analysis
we shall perform in the next section focuses on each subspace of
the phase space solutions separately. This means that the de
Sitter case $m=0$ has to be dealt separately from the $m=-3/2$
matter dominated epoch, so for each case the constraints found for
the parameters $\mu$ and $\lambda$ are different in the two cases.
Unfortunately there is no way to connect these eras in a physical
or mathematical way with the present framework, although in
general the couplings may change in an effective way, thus
connecting the various cosmological eras, and during each era
these may be approximately constant. But as we stated, we do not
have this theory at hand, so each study performed in the
following, has to be dealt separately from the other studies, and
there is no consistent mathematical or physical way in the
literature, to our knowledge, that enables one to connect these
cosmological eras.


\section{The Phase Space of the Model: Analytical Results}

Any autonomous dynamical system of the form,
\begin{equation*}
x^{\prime} = \mathcal{F}(x) \, ,
\end{equation*}
contains a number of invariant structures on the phase space, such
that its behavior is determined by them. These can be traced if we
consider that the vector field $\mathcal{F}(x)$ is zero or remains
constant on them. The first condition, that is $\mathcal{F}(x) =
0$, usually reveals the equilibrium points of the system, while
the second condition, namely $\nabla \mathcal{F}(x) = 0$, reveals
limit cycles or other attracting/repelling limit sets, such as for
example the stable or unstable manifolds of an equilibrium point.
After locating these invariant structures on the phase space of
the dynamical system, we may proceed by characterizing their
stability, usually by means of the Hartman-Grobman theorem and the
linearization of the vector field in small areas around them.

Setting $\mathcal{F}(x) = 0$ in our case, we may easily recover
four equilibrium points,
\begin{equation}\label{eq:Equilibria}
\begin{split}
&P_{0} \left( 0 , 0 , 0 , 0 , 0 \right) \; \, , \;\;\; P_{1} \left( -\dfrac{\sqrt{\frac{2}{3}} m}{\mu} , 0 , \dfrac{m(m+3)}{12 \mu (m+1)} , 0 , 0 \right) \; \, , \\
&P_{2} \left( -\dfrac{\sqrt{6} m}{\lambda} , -\dfrac{\sqrt{-2m^2
-6m}}{\lambda} , 0 , 0 , 0 \right) \;\;\; \text{and} \;\;\; P_{3}
\left( -\dfrac{\sqrt{6} m}{\lambda} , \dfrac{\sqrt{-2m^2
-6m}}{\lambda} , 0 , 0 , 0 \right) \, .
\end{split}
\end{equation}
We observe that all four points collide to $P_{0}$, as $m = 0$.
Checking if these four points fulfill the Friedmann constraint of
Eq. (\ref{eq:FriedConstraint}), we easily obtain the following
conclusions:
\begin{enumerate}

\item The point $P_{0}$ is easily proved to not fulfill the
constraint, thus it is deemed non-physical.

\item The point $P_{1}$ could fulfill the constraint, if $m \neq
0$ (so it exists apart from $P_{0}$) and
\begin{equation*}
\mu = \pm \dfrac{\sqrt{-2m^2 (m+5)}}{\sqrt{3(m+1)}} \, ;
\end{equation*}
this however indicates an imaginary $\mu$ for any $m<0$. Thus, the
point $P_{1}$ cannot be physical. A hint for determining the
importance and viability of this equilibrium point is the fact
that $x_{2}^{*} = 0$, essentially yielding a zero potential,
deeming thus the scalar field to be constant, and $x_{3}^{*} < 0$
for all $m<0$, which cannot occur due to our assumptions on the
form of $h(\phi)$.

\item Either the point $P_{2}$ or the point $P_{3}$ shall fulfill
the constraint. Both cannot be physical at the same time however.
If $\lambda = -\sqrt{2m(2 m-3)}$, then equilibrium $P_{2}$ is
physical, whereas if $\lambda = \sqrt{2m(2 m-3)}$, then
equilibrium $P_{3}$ is physical. Choosing the latter, so that
$\lambda > 0$ for $-3 \leq m<0$, we easily understand that the
point $P_{3}$ is the sole physical equilibrium of the system, much
more only when $m<0$ and $\lambda$ is constrained by $m$. The
viability of both (unphysical) $P_{2}$ and (physical) $P_{3}$ is
intriguing, since $x_{3}^{*} = 0$, essentially demanding that
either the coupling of the scalar field to the Gauss-Bonnet term
is constant, and thus $\mu = 0$, or that the Hubble rate tends to
zero, for all $-1<m<0$. Given the forms $H(t) \sim t^{-1}$ that we
will assume, the latter is indeed possible, allowing us a certain
freedom concerning $\mu$, and thus the latter shall be regarded as
the one and only true free parameter of the system.

\end{enumerate}
Proceeding with the linearization of the vector field in the
vicinity of each fixed point, we may examine their stability.
\begin{enumerate}
\item Concerning the point $P_{0}$, the linearized system takes
the form
\begin{equation}
\begin{pmatrix}
\xi_{1}^{\prime} \\ \xi_{2}^{\prime} \\ \xi_{3}^{\prime} \\
\xi_{4}^{\prime} \\ \xi_{5}^{\prime}
\end{pmatrix}
 =
\begin{pmatrix}
 -3-m & 0 & -4 \sqrt{6}(1+m) & 0 & 0 \\
 0 & -m & 0 & 0 & 0 \\
 0 & 0 & 2 m & 0 & 0 \\
 0 & 0 & 0 & -2-m & 0 \\
 0 & 0 & 0 & 0 & -\dfrac{3}{2}-m \\
\end{pmatrix}
\begin{pmatrix}
\xi_{1} \\ \xi_{2} \\ \xi_{3} \\ \xi_{4} \\ \xi_{5}
\end{pmatrix} \, ,
\end{equation}
where $\{ \xi_{i} \}$ are small linear perturbations of the
dynamical variables around the equilibrium point. The eigenvalues
of the system are,
\begin{equation*}
l_{1} = -(3+m) \; \, , \;\;\; l_{2} = -(2+m) \; \, , \;\;\; l_{3}
= -\Big(\dfrac{3}{2}+m\Big) \; \, , \;\;\; l_{4} = -m \;\;\;
\text{and} \;\;\; l_{5} = 2m \, .
\end{equation*}
We observe that, for $-3 \leq m \leq 0$, all eigenvalues are real,
but are subject to a bifurcation. $l_{1}$ is zero for $m=-3$ and
turns negative for $-3<m \leq 0$, deeming the direction  $v_{1} =
\vec{e}_{1} = (1,0,0,0,0)$ initially neutral and attractive
afterwards. In the same way $l_{2}$ turns from positive for $-3
\leq m < -2$ to negative for $-2 < m \leq 0$, thus the manifolds
tangent to $v_{2} = \vec{e}_{4} = (0,0,0,1,0)$ turns from unstable
to stable. The same occurs for $l_{1}$ in the case $m =
-\dfrac{3}{2}$, so that the manifolds tangent to $v_{3} =
\vec{e}_{5} = (0,0,0,0,1)$ is initially unstable and shifts to
stable when $m < -\dfrac{3}{2}$.\footnote{Notice that manifolds
$v_{2}$ and $v_{3}$ are centre manifolds for $m = -2$
(radiation-dominated Universe) and $m = -\dfrac{3}{2}$
(matter-dominated Universe) respectively. This result, occurring
at each point, is foreshadowed in section III.A and indicates the
constancy (or neutrality) of each of the density parameters in its
corresponding domination.}  Furthermore, the point is subject to
two bifurcations for $m = 0$. In the case where $m = 0$ (de Sitter
expansion), $l_{4} = l_{5} = 0$, so their respective manifolds are
centre manifolds, and thus slow or null evolution of the
respective variables occurs on them. When $m < 0$, then the
manifold tangent to $v_{4} = \Big( 4\sqrt{\dfrac{2}{3}},0,1,0,0
\Big)$ becomes stable, while the manifold tangent to $v_{5} =
\vec{e}_{2} = (0,1,0,0,0)$ becomes unstable.\footnote{Notice that
for $m>0$, the stability of the two manifolds is reversed.} Thus,
the overall behavior of the equilibrium point $P_{0}$ is that of a
saddle with attractive and repulsive directions shifting along the
values of $m$.

\item Concerning the point $P_{1}$, the linearized system takes
the form,
\begin{equation}
\begin{pmatrix}
\xi_{1}^{\prime} \\ \xi_{2}^{\prime} \\ \xi_{3}^{\prime} \\
\xi_{4}^{\prime} \\ \xi_{5}^{\prime}
\end{pmatrix}
 =
\begin{pmatrix}
 -3-m & 0 & -4 \sqrt{6}(m+1) & 0 & 0 \\
 0 & \dfrac{m \lambda}{3 \mu}-m & 0 & 0 & 0 \\
 \dfrac{m(m+3)}{2\sqrt{6} (m+1)} & 0 & 0 & 0 & 0 \\
 0 & 0 & 0 & -2-m & 0 \\
 0 & 0 & 0 & 0 & -\dfrac{3}{2}-m \\
\end{pmatrix}
\begin{pmatrix}
\xi_{1} \\ \xi_{2} \\ \xi_{3} \\ \xi_{4} \\ \xi_{5}
\end{pmatrix} \, .
\end{equation}
The eigenvalues of the system are,
\begin{align*}
l_{1} &= -(2+m) \; \, , \;\;\; l_{2} = -\Big(\dfrac{3}{2}+m\Big) \; \, , \;\;\; l_{3} = \dfrac{m (\lambda -3\mu)}{3\mu} \; \, , \\
l_{4} &= \dfrac{-3 -4m -m^2 - \mathcal{S}}{2(m+1)} \;\;\;
\text{and} \;\;\; l_{5} = \dfrac{-3 -4m -m^2 +
\mathcal{S}}{2(m+1)} \; \, ,
\end{align*}
where $\mathcal{S} = \sqrt{9 -34m^2 -32m^3 -7m^4}$. As in $P_{0}$,
for $-3 \leq m \leq 0$, two eigenvalues ($l_{1}$ and $l_{2}$)
shift from positive to negative at $m = -2$ and $m =
-\dfrac{3}{2}$ respectively, thus the manifolds tangent to the
directions $v_{1} = \vec{e}_{4} = (0,0,0,1,0)$ and $v_{2} =
\vec{e}_{5} = (0,0,0,0,1)$ are initially repulsive and eventually
attractive. Eigenvalue $l_{3}$ is subject to two bifurcations, one
for $m = 0$ and one for $\lambda = 3\mu$.\footnote{Notice also the
singularity at $\mu = 0$, a value of little practical use so far.}
In the first case, when $m = 0$, the direction of $v_{3} =
\vec{e}_{2} = (0,1,0,0,0)$ is neutral, while for $m < 0$ and
$\lambda > 3\mu$ it is attractive. However, when $\lambda = 3\mu$,
this manifolds becomes neutral again, and when $\lambda < 3\mu$
for $m < 0$, the manifold tangent to $v_{3}$ becomes unstable.
Finally, as long as eigenvalues $l_{4}$ and $l_{5}$ are concerned,
both are proved to be real but of different sign in the intervals
$-1<m<0$ and $-3<m<-1$, for $\mu > 0$; we notice that $l_{4} =
l_{5} = 0$ for $m = 0$ (de Sitter expansion) and $m = -3$ (stiff
matter-dominated Universe). Subsequently, the corresponding
eigenvectors,
\begin{align*}
v_{4} &= \Bigg( \dfrac{\sqrt{6} \left(-3 -4m -m^2 +\sqrt{-(m+1)^2 \left( 7m^2 +18m -9 \right)}\right)}{m^2 +3m} , 0 , 1 , 0 , 0 \Bigg) \;\;\; \text{and} \\
v_{5} &= \Bigg( -\dfrac{\sqrt{6} \left(-3 -4m -m^2 +\sqrt{-(m+1)^2
\left( 7m^2 +18m -9 \right)}\right)}{m^2 +3m} , 0 , 1 , 0 , 0
\Bigg) \, ,
\end{align*}
denote an unstable and a stable manifold respectively for $-1<m<0$
and the opposite for $-3<m<-1$, and two centre manifolds for $m =
0$ and $m = -3$.

\item Concerning the point $P_{2}$, the linearized system takes
the form,
\begin{equation}
\begin{pmatrix}
\xi_{1}^{\prime} \\ \xi_{2}^{\prime} \\ \xi_{3}^{\prime} \\
\xi_{4}^{\prime} \\ \xi_{5}^{\prime}
\end{pmatrix}
 =
\begin{pmatrix}
 -3-m & -2 \sqrt{-3(m^2 +3m)} & -4 \sqrt{6}(m+1) & 0 & 0 \\
 \sqrt{\dfrac{-m^2 -3m}{3}} & 0 & 0 & 0 & 0 \\
 0 & 0 & 2 \left(m - \dfrac{3m \mu}{\lambda}\right) & 0 & 0 \\
 0 & 0 & 0 & -2-m & 0 \\
 0 & 0 & 0 & 0 & -\dfrac{3}{2}-m \\
\end{pmatrix}
\begin{pmatrix}
\xi_{1} \\ \xi_{2} \\ \xi_{3} \\ \xi_{4} \\ \xi_{5}
\end{pmatrix} \, ,
\end{equation}
and yields the following eigenvalues,
\begin{align*}
l_{1} &= -(2+m) \; \, , \;\;\; l_{2} = -\Big(\dfrac{3}{2}+m\Big) \; \, , \;\;\; l_{3} = \dfrac{2m (\lambda -3\mu)}{\lambda} \; \, , \\
l_{4} &= \dfrac{-3 -m -\sqrt{3 \left(3 +10m +3m^2\right)}}{2}
\;\;\; \text{and} \;\;\; l_{5} = \dfrac{-3 -m +\sqrt{3 \left(3
+10m +3m^2\right)}}{2} \; \, .
\end{align*}
Again, for $-3 \leq m \leq 0$, two eigenvalues ($l_{1}$ and
$l_{2}$) shift from positive to negative at $m = -2$ and $m =
-\dfrac{3}{2}$ respectively, thus the manifolds tangent to the
directions $v_{1} = \vec{e}_{4} = (0,0,0,1,0)$ and $v_{2} =
\vec{e}_{5} = (0,0,0,0,1)$ are initially repulsive and eventually
attractive. The third eigenvalue is zero when $m = 0$, or when $m
< 0$ and $\lambda = 3\mu$. This corresponds to a neutrality of the
manifold tangent to,
\begin{equation*}
v_{3} = \Bigg( -\dfrac{4\sqrt{6}\lambda \left( \lambda-3\mu
\right)(1+m)}{-9\lambda\mu +2\lambda^2 m -15\lambda\mu m +18\mu^2
m} , -\dfrac{2\lambda^2 (1+m) \sqrt{-2m(m+3)}}{m
\left(-9\lambda\mu +2\lambda^2 m -15\lambda\mu m +18\mu^2
m\right)} , 1 , 0 , 0 \Bigg) \, ;
\end{equation*}
this very manifold turns from stable when $\lambda > 3\mu$ to
unstable when $\lambda < 3\mu$.\footnote{Notice the singularity
reached as $\lambda = 0$, that has little to no practical
importance in our case.} As for the remaining two eigenvalues,
these also contain a bifurcation: they are real and negative for
$-\dfrac{1}{3} \leq m \leq 0$, complex with negative real parts
for $-3 < m < -\dfrac{1}{3}$ and zero for $m = -3$. Thus the
corresponding eigenvectors,
\begin{align*}
v_{4} &= \Bigg( \dfrac{\sqrt{3} \left(-3 -m +\sqrt{3\left(3 m^2+10 m+3\right)}\right)}{2 \sqrt{-m(m+3)}} , 1 , 0 , 0 , 0 \Bigg) \; \, , \\
v_{5} &= \Bigg( -\dfrac{\sqrt{3} \left(-3 -m +\sqrt{3\left(3
m^2+10 m+3\right)}\right)}{2 \sqrt{-m(m+3)}} , 1 , 0 , 0 , 0
\Bigg)
\end{align*}
denote the presence of stable manifolds, either in direct or in
oscillatory attraction towards equilibrium $P_{2}$, and a neutral
behavior in the case of $m = -3$ (stiff matter-dominated
Universe).

\item Finally, in the area of equilibrium point $P_{3}$, the
system is linearized as,
\begin{equation}
\begin{pmatrix}
\xi_{1}^{\prime} \\ \xi_{2}^{\prime} \\ \xi_{3}^{\prime} \\
\xi_{4}^{\prime} \\ \xi_{5}^{\prime}
\end{pmatrix}
 =
\begin{pmatrix}
 -3-m & -2 \sqrt{-3(m^2 +3m)} & -4 \sqrt{6}(m+1) & 0 & 0 \\
 \sqrt{\dfrac{-m^2 -3m}{3}} & 0 & 0 & 0 & 0 \\
 0 & 0 & 2 \left(m - \dfrac{3m \mu}{\lambda}\right) & 0 & 0 \\
 0 & 0 & 0 & -2-m & 0 \\
 0 & 0 & 0 & 0 & -\dfrac{3}{2}-m \\
\end{pmatrix}
\begin{pmatrix}
\xi_{1} \\ \xi_{2} \\ \xi_{3} \\ \xi_{4} \\ \xi_{5}
\end{pmatrix} \, ,
\end{equation}
and yields the following eigenvalues,
\begin{align*}
l_{1} &= -(2+m) \; \, , \;\;\; l_{2} = -\Big(\dfrac{3}{2}+m\Big) \; \, , \;\;\; l_{3} = \dfrac{2m (\lambda -3\mu)}{\lambda} \; \, , \\
l_{4} &= \dfrac{-3 -m -\sqrt{3 \left(3 +10m +3m^2\right)}}{2}
\;\;\; \text{and} \;\;\; l_{5} = \dfrac{-3 -m +\sqrt{3 \left(3
+10m +3m^2\right)}}{2} \; \, .
\end{align*}
It is obvious that whatever said concerning the stability of point
$P_{2}$ holds equally for point $P_{3}$. Those two are symmetric
over the $x_{2}$ axis, sharing the same stability properties.

\end{enumerate}
From this analysis, it is clear that equilibrium point $P_{0}$ may
have up to four attractive directions with two centre manifolds
appearing for $m = 0$, one for $m = -\dfrac{3}{2}$, one for $m =
-2$. This fact, along with its non-viability, mark equilibrium
$P_{0}$ as an improbable resolution for the system, while the only
case of physical importance in which it is fully hyperbolic is
when $m = -1$. In the same manner, equilibrium point $P_{1}$ has
at least one unstable manifold and exhibits centre manifolds for
$m = 0$, one for $m = -\dfrac{3}{2}$, one for $m = -2$ and one for
$m = -3$. Another unphysical point, with a subsequent singularity
for $m = -1$, cannot be counted as the relaxation point of our
system. Finally, equilibria $P_{2}$ and $P_{3}$ are symmetric and
under circumstances are globally stable, with all five manifolds
being attractive. Such a case is when $m = -1$, however centre
manifolds appear for it as well in the physically interesting
cases of $m = 0$, $m = -\dfrac{3}{2}$, $m = -2$ and $m = -3$.

Discarding the behavior of $x_{4}$ and $x_{5}$ as trivial, we are
immediately spared the bifurcations of $m = -\dfrac{3}{2}$ and $m
=2$, hence the behavior of the system of Eqs. (\ref{eq:firstODE}),
(\ref{eq:secondODE}) and (\ref{eq:thirdODE}) is quite normal,
proposing that $\lambda > 3\mu$ ( that is for $\lambda \gg \mu$,
instead of $\lambda = \mu$, which we shall both specifically
examine) and $\lambda = \sqrt{2m(2m-3)}$ (to ensure the viability
of $P_{3}$). Furthermore, we are able to concentrate on the
remaining cases. What we notice is that all other interesting
phases of the Universe, stated in section III.C (notably the de
Sitter expansion and the stiff matter domination), are cases of
bifurcations, where the equilibrium points are non-hyperbolic.
Consequently, the behavior of the system cannot be described by
the linearization we conducted so far and the need for numerical
solutions to reveal the attractor behavior.
\\

Another important feature of the phase space arises when we set
$\nabla \mathcal{F} = 0$ and solve with respect to the dynamical
variables. It is easy to see that,
\begin{equation*}
\nabla \mathcal{F} = 2 \left( m+\sqrt{\dfrac{3}{2}}\mu x_{1}
\right) -4m -\dfrac{\lambda x_{1}}{\sqrt{6}}-\frac{13}{2}
\end{equation*}
is zero when $\lambda \neq 6\mu$ and,
\begin{equation}\label{eq:Repel1}
x_{1} = -\sqrt{\dfrac{3}{2}} \dfrac{(4m+13)}{\lambda -6\mu} \, .
\end{equation}
This hypersurface is proved to be an invariant in the phase space
when,
\begin{equation}\label{eq:Repel2}
x_{3} = \dfrac{4 m^2 +25m + 39 +\lambda x_{2}^2 (\lambda
-6\mu)}{8(m+1) (\lambda -6\mu)} \, .
\end{equation}
Eqs. (\ref{eq:Repel1}) and (\ref{eq:Repel2}) define an invariant
curve in the phase space, one that acts either as a repeller
($\alpha$-limit set) or an attractor ($\omega$-limit set) for the
system of Eqs. (\ref{eq:firstODE}),(\ref{eq:secondODE}) and
(\ref{eq:thirdODE}). It can be proved, rather easily by means of
linear perturbations, that it is in fact a repeller, so its
practical importance is limited. Furthermore, given $\lambda =
6\mu$, both $x_{1}$ and $x_{3}$ tend to infinity, thus the curve
is removed from the phase space. One more important element is
that, for $-1<m\leq0$, both $x_{1}$ and $x_{3}$ are negative, so
they have no physical meaning whatsoever.

Another interesting feature of the dynamical system is the
existence of a curve defined as,
\begin{equation}\label{eq:Attract}
C: \; x_{1} = -\dfrac{\sqrt{6} m}{\lambda} \;\;\; \text{and}
\;\;\; x_{3} = \dfrac{2m^2 +6m +\lambda^2 x_{2}^2}{8\lambda(m+1)}
\, .
\end{equation}
The action of the vector field $\mathcal{F}(x)$ on this curve, is,
\begin{equation}
\mathcal{F}_{1} \dfrac{\partial C}{\partial x_{1}} +
\mathcal{F}_{2} \dfrac{\partial C}{\partial x_{2}} +
\mathcal{F}_{3} \dfrac{\partial C}{\partial x_{3}} \, ,
\end{equation}
and is identically zero if $\lambda = 3\mu$ or $m = 0$, which are
the values of the parameters on which bifurcations occur. In other
words, we have traced the shifting manifold of equilibrium points
$P_{2}$ and $P_{3}$ -which turn to $P_{0}$ for $m = 0$. This
manifold has been identified as a neutral for $m=0$ or $\lambda =
3\mu$, so it functions as a locus of equilibria in these cases.

We may easily apply linear perturbation theory on this curve, as
well, leading to the linearized system,
\begin{equation}
\begin{pmatrix}
\psi_{1}^{\prime} \\ \psi_{2}^{\prime} \\ \psi_{3}^{\prime} \\
\psi_{4}^{\prime} \\ \psi_{5}^{\prime}
\end{pmatrix}
 =
\begin{pmatrix}
 -m-3 & \sqrt{6} \lambda x_{2} & -4 \sqrt{6}(1+m) & 0 & 0 \\
 -\dfrac{\lambda}{\sqrt{6}} x_{2} & 0 & 0 & 0 & 0 \\
 \sqrt{\dfrac{3}{2}} \dfrac{\mu\left(2m^2 +6m + \lambda^2 x_{2}^2 \right)}{4(m+1)\lambda } & 0 & 2 \left( m -\dfrac{3m \mu}{\lambda} \right) & 0 & 0 \\
 0 & 0 & 0 & -2-m & 0 \\
 0 & 0 & 0 & 0 & -\dfrac{3}{2}-m \\
\end{pmatrix}
\begin{pmatrix}
\psi_{1} \\ \psi_{2} \\ \psi_{3} \\ \psi_{4} \\ \psi_{5}
\end{pmatrix} \, ,
\end{equation}
where $\{ \psi_{i} \}$ are small linear perturbations of the
dynamical variables around the curve $C$. Notice that these
perturbations, unlike perturbations around an equilibrium point,
dependent on the variable $x_{2}$, which may be treated as a free
parameter so long as it is restricted on the curve $C$. The
eigenvalues of the system are,
\begin{equation*}
l_{1} = -(2+m) \; \, , \;\;\; l_{2} = -\Big(\dfrac{3}{2}+m\Big) \;
\, , \;\;\; l_{3} = \dfrac{\mathcal{R}_{1}}{24(1+m)\lambda} \; \,
, \;\;\; l_{4} = \dfrac{\mathcal{R}_{2}}{24(1+m)\lambda} \;\;\;
\text{and} \;\;\; l_{5} = \dfrac{\mathcal{R}_{3}}{24(1+m)\lambda}
\, ,
\end{equation*}
where $\mathcal{R}_{1}$, $\mathcal{R}_{2}$ and $\mathcal{R}_{3}$
are the roots of polynomial
\begin{equation*}
\begin{split}
&-27648m (m+1)^3 \lambda^4 \left(\lambda -3\mu \right) x_{2}^2 -576(m+1)^2 \lambda \left( 2m^2 (\lambda -6\mu) +6m (\lambda -6\mu)-\lambda^2 (\lambda +3\mu) x_{2}^2 \right) \mathcal{R} -  \\
&-24(m+1)\left(\lambda(m-3) -6m \mu \right) \mathcal{R}^2 +
\mathcal{R}^3 = 0 \, ,
\end{split}
\end{equation*}
the first of which is real and negative, while the other two are
complex with negative real parts, in the regime of our interest
($-1< m \leq 0$ and $\lambda > 6\mu$) and for physical values of
the potential ($ 0 \geq x_{2} > 10$). As a result, it is easy to
conclude that the curve $C$ is an attractive invariant of the
system, that offers a multiplicity of stationary solutions in the
abnormal cases of $m = 0$ and $\lambda = 3\mu$.

Aside from the stability of these manifolds, we may also deal with
their viability. Setting $x_{1} = -\dfrac{\sqrt{6} m}{\lambda}$,
$x_{3} = \dfrac{2m^2 +6m +\lambda^2 x_{2}^2}{8\lambda(m+1)}$ and
$x_{4} = x_{5} = 0$ in Eq. (\ref{eq:FriedConstraint}), we see
that,
\begin{equation*}
x_{1}^2 + x_{2}^2 + 8\sqrt{6} x_{1} x_{3} + x_{4}^2 + x_{5}^2 =
\dfrac{\lambda^2 (1-5m) x_{2}^2 -6 m^2(m+5)}{\lambda^2(m+1)} \, ,
\end{equation*}
that can equal unity if and only if,
\begin{equation}
x_{2}^2 = -\dfrac{6m^2 (m+5) +(m+1)\lambda^2}{(5m-1)\lambda^2} \,
.
\end{equation}
Essentially, the system may always end up on the invariant curve
$C$, when surpassing a bifurcation -when $m=0$ or $\lambda =
3\mu$- but its relaxation is not always physical. In order for the
Friedmann constraint to be fulfilled and the viability to be
ensured, special initial conditions are needed (a short of
fine-tuning) that may lead to specific values of $x_{2}$. As we
notice, these values depend on $m$ and $\lambda$, letting $\mu$ as
a free parameter. The physical meaning of these is the following:
in the de Sitter case of the Universe, or in the case where the
potential decreases trice as fast as the coupling increases, with
respect to the scalar field, the potential must reach a specific
non-zero value so that the Universe will end up on the physical
stable de Sitter attractor.

The analysis we performed above is quite rigorous and
mathematically rigid, however we need to enlighten the physics of
the phase space with numerical analysis. In this way we may
visualize all the invariant manifolds and curves we discovered in
this section. This is compelling for a more persuasive treatment
of the problem, and it is the subject of the next section.


\section{The Phase Space of the Model: Numerical solutions}

In order to better understand the phase space structure of the
Einstein-Gauss-Bonnet models, we shall numerically integrate the
differential equations (\ref{eq:firstODE}), (\ref{eq:secondODE})
and (\ref{eq:thirdODE}) for a number of cases with physical
interest, as stated above. The integration of Eqs.
(\ref{eq:fourthODE}) and (\ref{eq:fifthODE}) is conducted
analytically, so they will not trouble us. The Universe should not
be considered vacuum, with the presence of matter fields being
mimicked by the modification of gravity along with the specific
choice of the Hubble rate, and thus of the parameter $m$, as
usually; this would lead to the misperception of certain
solutions. We need to keep in mind that only the reduced phase
space is studied here, hence specific trajectories that appear to
be unviable could be viable -and the vice versa. Whenever
necessary, we shall comment on the behaviour of $x_{4}$ and
$x_{5}$, though this behavior depends largely on the Hubble rate,
rather than that of the mass-energy densities.

For a specific case, the integration of the ``inverse problem''
($x^{\prime} = -\mathcal{F}(x)$) was also utilized, in order to
find special initial conditions that are needed to realize a
specific behaviour.

\subsection{de Sitter expanding Universe: $m = 0$}

Setting $m = 0$, then Eqs. (\ref{eq:firstODE}),
(\ref{eq:secondODE}) and (\ref{eq:thirdODE}) reduce to,
\begin{align}
x_{1}^{\prime} &= - 3x_{1} + \dfrac{\sqrt{6}}{2}\lambda x_{2}^2 - 4\sqrt{6} x_{3} \; \, , \\
x_{2}^{\prime} &= -\dfrac{\sqrt{6}}{2} \lambda x_{1} x_{2} \;\; \text{and} \\
x_{3}^{\prime} &= \dfrac{\sqrt{6}}{2}\mu x_{1} x_{3} \, .
\end{align}
The above dynamical system has one equilibrium point, $P_{0}$,
that has been proved to possess one stable and two centre
manifolds. Furthermore, there exists the curve $C: \; x_{3} =
\dfrac{\lambda}{8} x_{2}^2$ for $x_{1} = 0$, which is an attractor
for the solutions of the system, one of the centre manifolds of
point $P_{0}$. The attraction towards curve $C$ occurs by means of
oscillations for $x_{2} > \dfrac{3}{2 \sqrt{\lambda^2 +3\lambda
\mu}}$, and by means of direct attraction for $x_{2} < \dfrac{3}{2
\sqrt{\lambda^2 +3\lambda \mu}}$.

\begin{figure}
\includegraphics[width=0.45\linewidth]{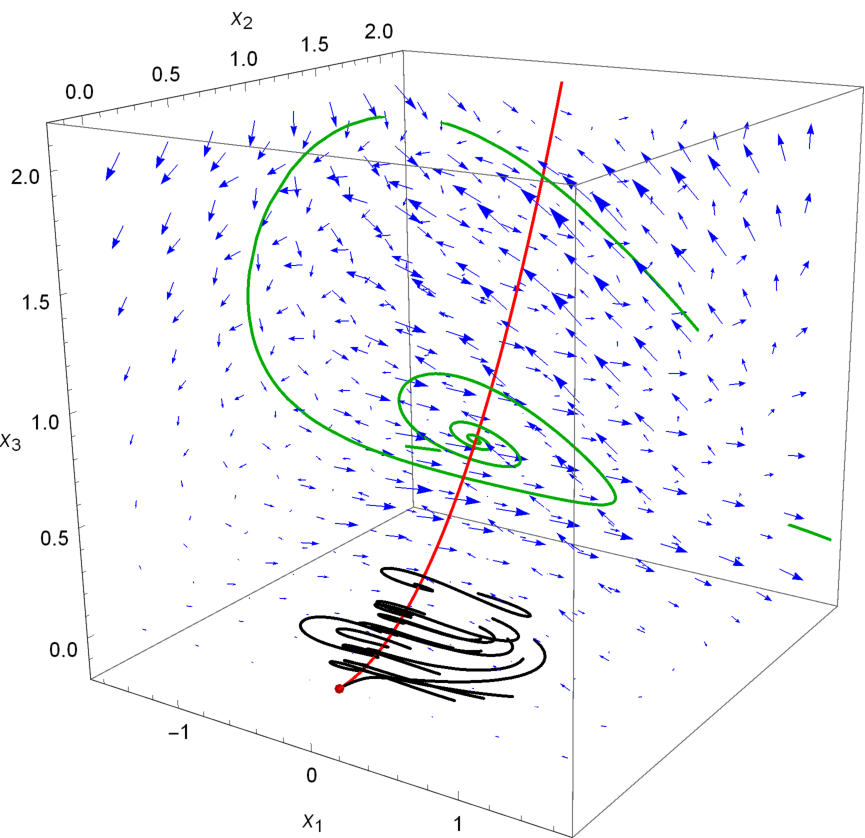}
\includegraphics[width=0.45\linewidth]{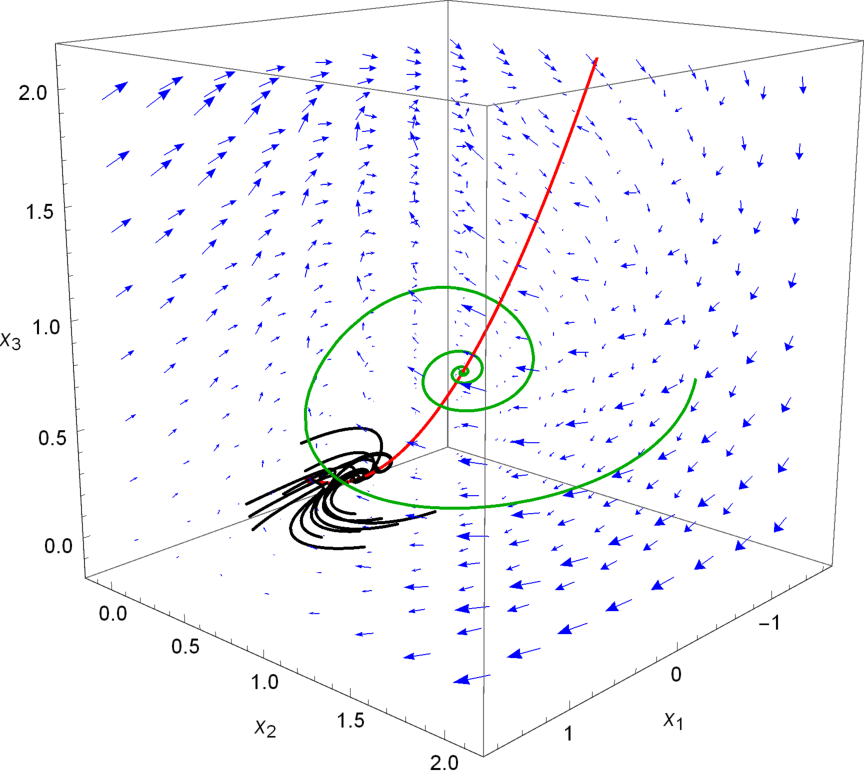}
\includegraphics[width=0.45\linewidth]{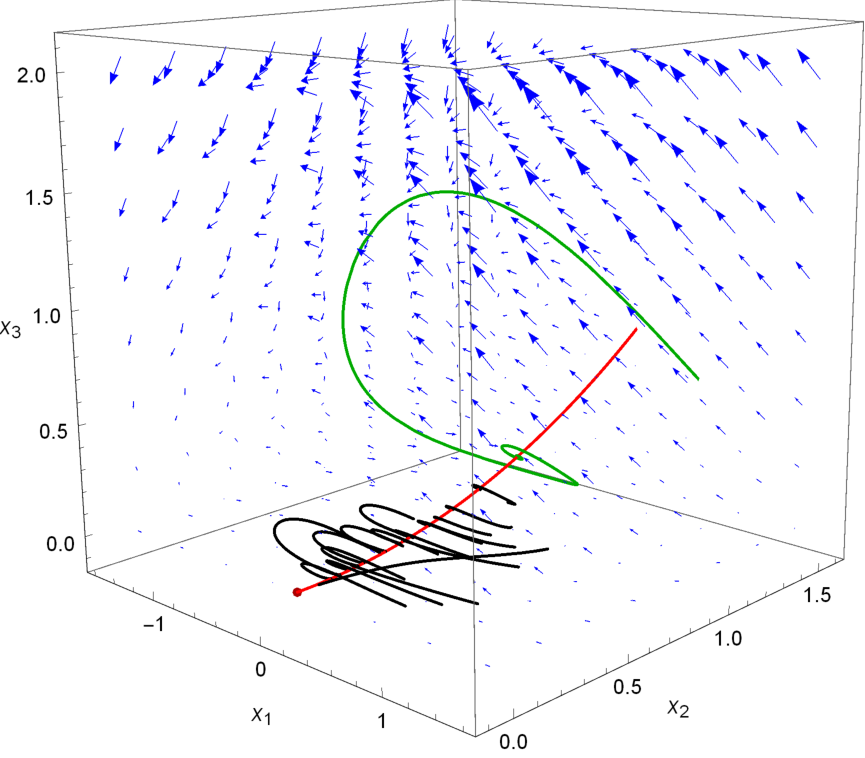}
\includegraphics[width=0.45\linewidth]{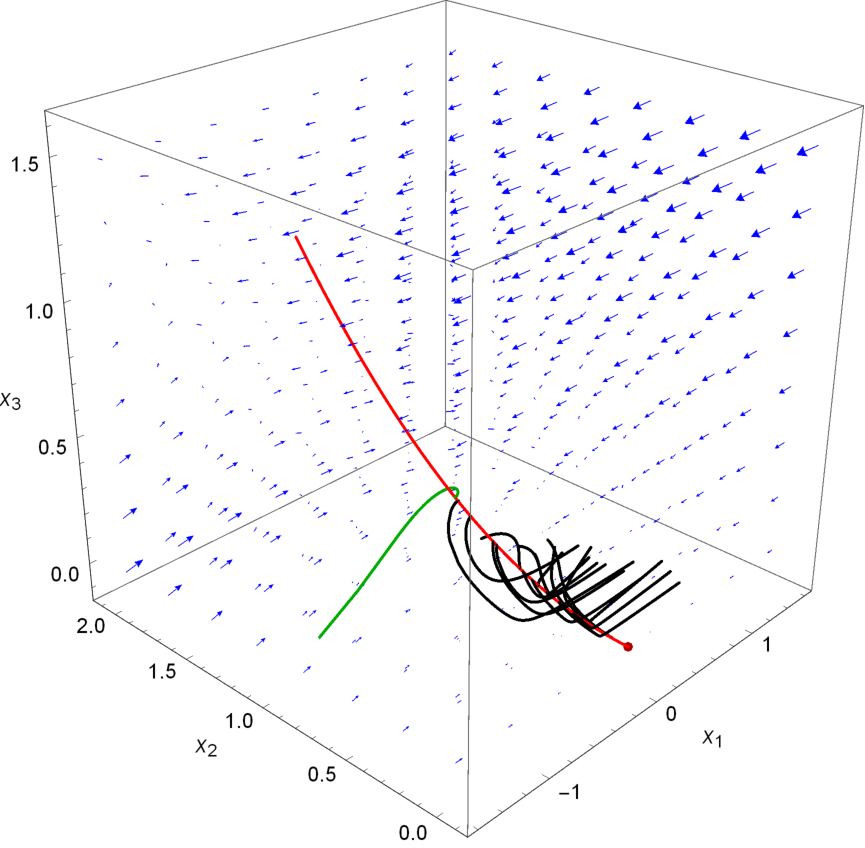}
\caption{The phase space of $x_1$, $x_2$ and $x_3$ for the de
Sitter expanding Universe, for $\lambda = 3\mu$, $\lambda = 6\mu$,
$\lambda = \mu$ and $\lambda \gg \mu$ (from top-left to
bottom-right). Blue arrows denote the vector field, black lines
denote numerical solutions of the system; the red curve stands for
the centre manifold $x_{3} = \dfrac{\lambda}{8} x_{2}^2$ and the
equilibrium point $P_{0}$ is marker dark red.}
\label{fig:deSitter_Numerics}
\end{figure}

What we actually expect is that all solutions of the system will
reach this attractor and remain fixed on some point of it. The
exact point is not important, except for $x_{1}^{*} = 0$,
$x_{2}^{*} = 1$ and $x_{3}^{*} = \dfrac{\lambda}{8}$, that
fulfills the Friedmann constraint. Of course, this is only one of
the infinite points on which the dynamical system may be attracted
to. In order to trace initial conditions that lead to the physical
point of curve $C$, we employed the technique of the ``inverse
problem'' and located one trajectory leading to this point -the
trajectory colored dark green in each subplot of Fig.
\ref{fig:deSitter_Numerics}. It is worth noting that, despite all
other trajectories -the black ones- have been derived from initial
conditions that respected the Friedmann constraint (Eq.
(\ref{eq:FriedConstraint}), this particular trajectory -and any
similar- initiates from sections of the phase space that do not
fulfill the constraint. This is an important feature to be
noticed.

In this case, $\lambda = \sqrt{2m(2m - 3)} = 0$ is of little
physical meaning (constant potential over the scalar field), thus
it will not be used. On the contrary, we chose four cases that
could be of interest: $\lambda = 3\mu$, a case of bifurcations for
$m \neq 0$; $\lambda = 6\mu$ and $\lambda = \mu$, before and after
the bifurcation, and $\lambda \gg \mu$, that is expected to have
some physical meaning (a potential evolving faster over $\phi$
than the coupling function $h(\phi)$ to the Gauss-Bonnet term).
All cases yielded similar results, with all the solutions reaching
normally the attraction curve $C$. Only in the last case, the
attractions were direct, instead of oscillatory, due to the
relationship between $\lambda$ and $\mu$.

During this era, both $x_{4}$ and $x_{5}$ are decreasing fast
towards their equilibrium values $x_{4}^{*} = x_{5}^{*} = 0$.

\subsection{Radiation-dominated Universe: $m = -2$}
\begin{figure}
\includegraphics[width=0.45\linewidth]{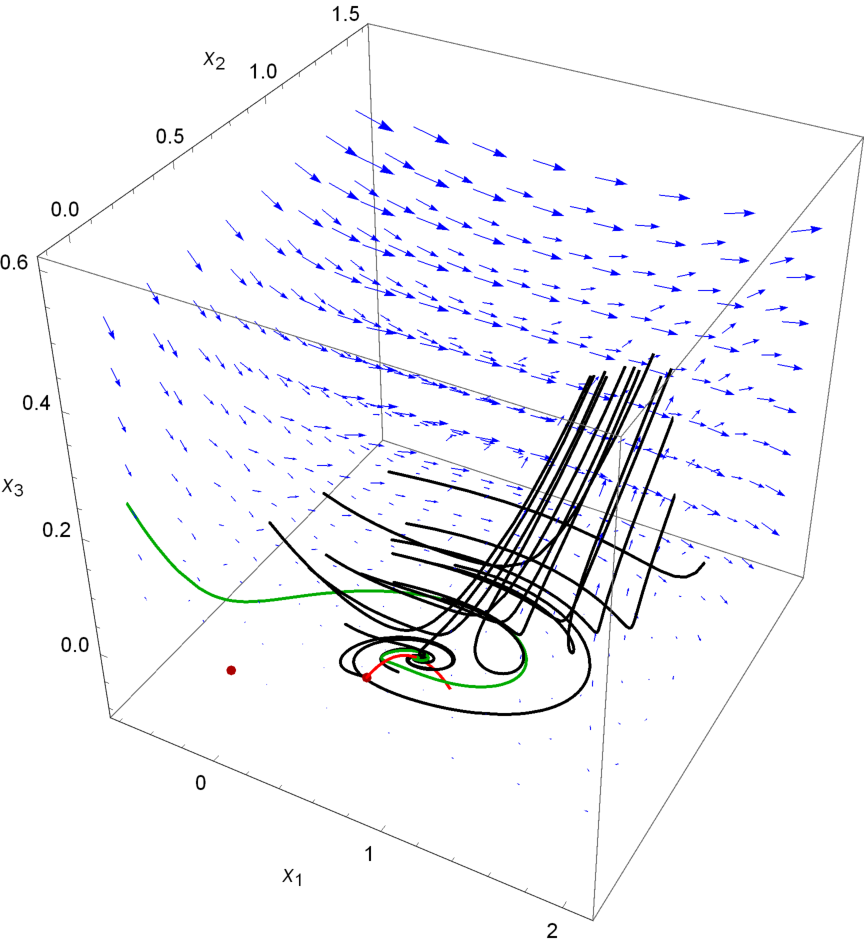}
\includegraphics[width=0.45\linewidth]{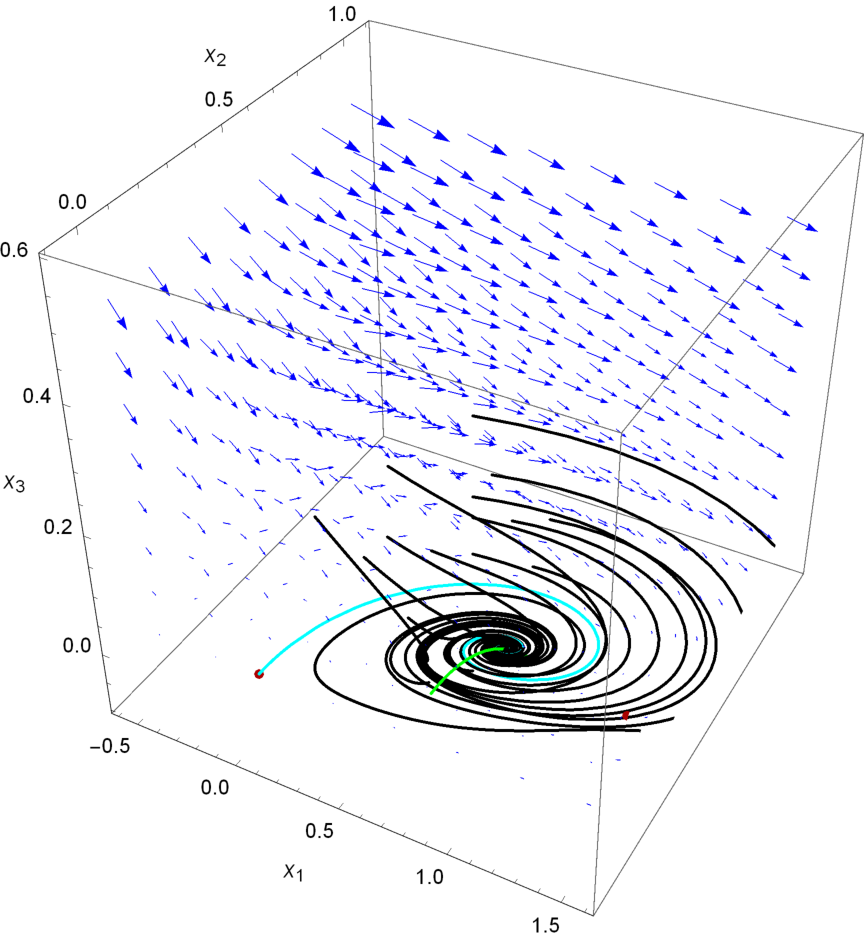}
\includegraphics[width=0.45\linewidth]{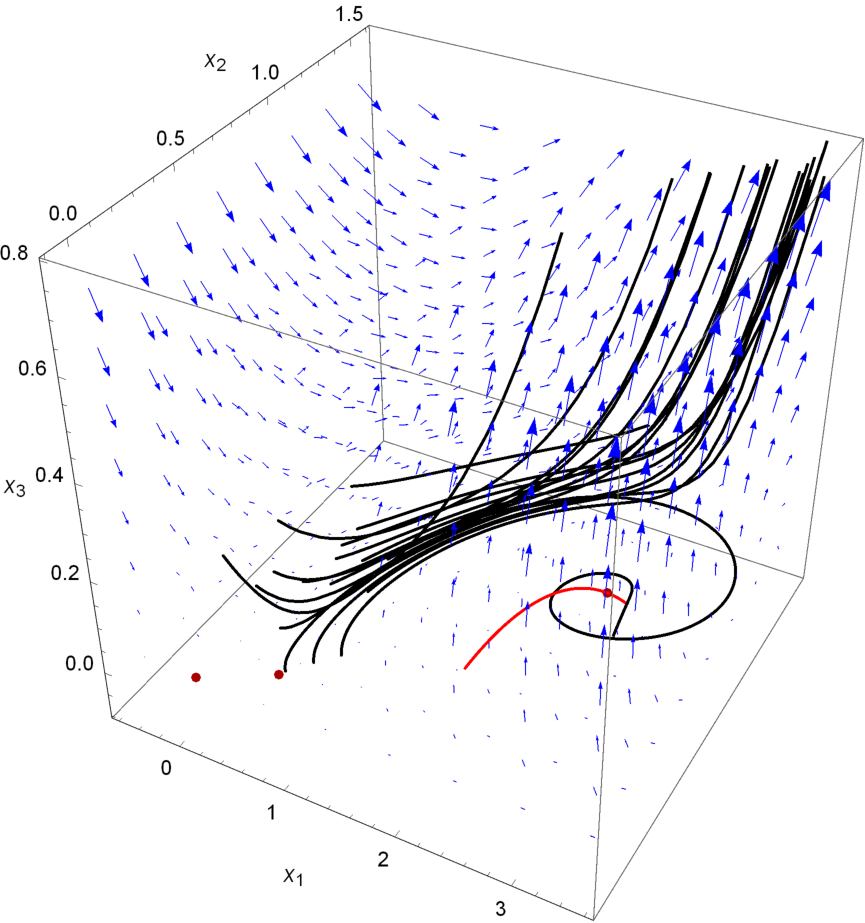}
\includegraphics[width=0.45\linewidth]{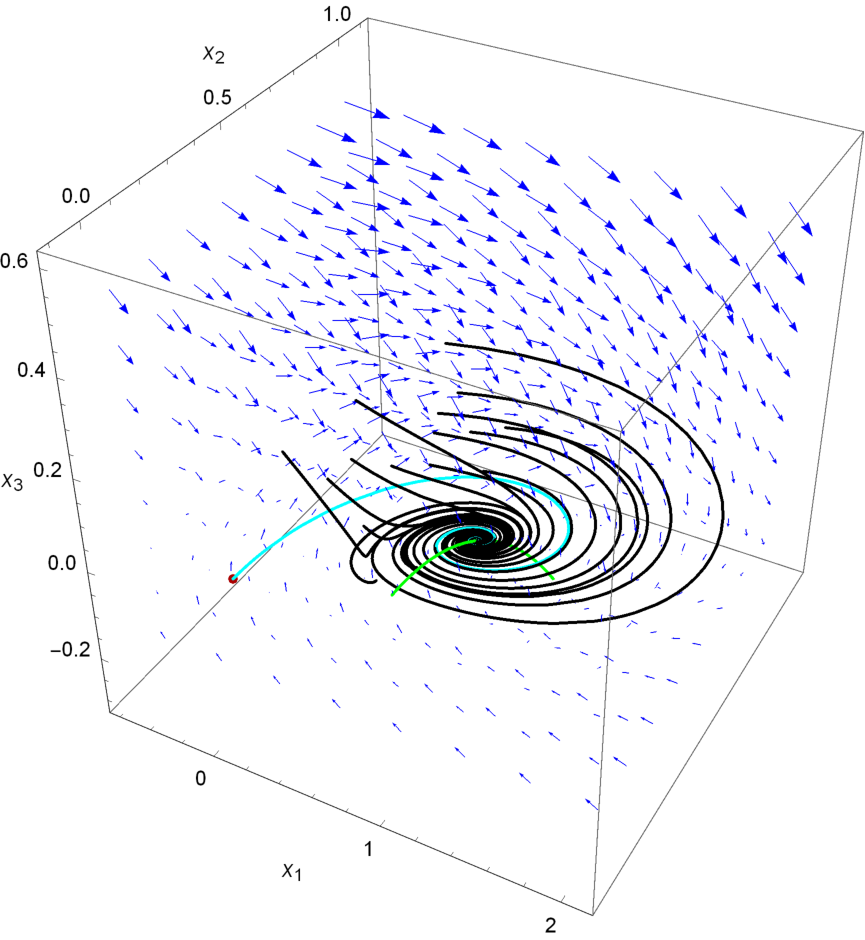}
\caption{The phase space of $x_1$, $x_2$ and $x_3$ for the
radiation-dominated Universe, for $\lambda = 3\mu$, $\lambda =
6\mu$, $\lambda = \mu$ and $\lambda \gg \mu$ (from top-left to
bottom-right). Blue arrows denote the vector field, black lines
denote numerical solutions of the system; the red curve stands for
the unstable manifold and the green curve for the stable manifold,
while the cyan curve depicts the heteroclinic curve from $P_{0}$
to $P_{3}$. The equilibrium points are marked dark red if they are
unviable and dark green if they are
viable.}\label{fig:RadiationNumerics}
\end{figure}
Setting $m = -2$, then Eqs. (\ref{eq:firstODE}),
(\ref{eq:secondODE}) and (\ref{eq:thirdODE}) reduce to,
\begin{align}
x_{1}^{\prime} &= - x_{1} + \dfrac{\sqrt{6}}{2}\lambda x_{2}^2 + 4\sqrt{6} x_{3} \; \, , \\
x_{2}^{\prime} &= -x_{2} \Bigg( \dfrac{\sqrt{6}}{2} \lambda x_{1} -2 \Bigg) \;\; \text{and} \\
x_{3}^{\prime} &= x_{3} \Bigg( \dfrac{\sqrt{6}}{2}\mu x_{1} -2
\Bigg) \, .
\end{align}
These equations exhibit all four equilibrium points we have
mentioned. Point $P_{0}$ possesses two stable and one unstable
manifolds. Point $P_{1}$ possesses one stable and one unstable
manifold, while the third, tangent to $\vec{e}_{2} = (0,1,0)$,
shifts from stable to unstable as $\lambda = 3\mu$; finally,
points $P_{2}$ and $P_{3}$ yield two complex eigenvalues with
negative real parts (so attraction with oscillations) and a third
real eigenvalue that shifts from negative to positive as $\lambda
= 3\mu$. As a result, we shall use the same values of $\lambda$ as
with the previous case ($m = 0$). Furthermore, in the last case
$\lambda = \sqrt{2m(2m-3)} = 2\sqrt{7}$ is chosen as to ensure the
viability of point $P_{3}$. We should notice that the curve $C$
takes the form,
\begin{equation}
x_{3} = \dfrac{4 -\lambda^2 x_{2}^2}{8\lambda}\, ,
\end{equation}
and acts either as stable, centre or unstable manifold for the
system, depending on the relationship between $\lambda$ and $\mu$.
The stable cases are denoted as green curves, whereas the unstable
ones are denoted as red in Fig. \ref{fig:RadiationNumerics}. We
also notice the existence of a heteroclinic curve leading from
equilibrium $P_{0}$ to equilibrium $P_{3}$. This curve is painted
cyan in two of the subplots of Fig. \ref{fig:RadiationNumerics}.

In the case of $\lambda = 3\mu$ (the top-left subplot of Fig.
\ref{fig:RadiationNumerics}), where the manifold is neutral, we
solved the ``inverse problem'' tracing a trajectory that reaches a
physical point in the phase space, one for which $x_{1} =
\dfrac{2\sqrt{6}}{\lambda}$ , $x_{2} = \dfrac{1}{\lambda}
\sqrt{\dfrac{24 + \lambda^2}{13}}$ and $x_{3} = \dfrac{28
-\lambda^2}{104 \lambda}$; this trajectory is marked dark green in
the first subplot of Fig. \ref{fig:RadiationNumerics}. Once more,
while all numerical solutions were derived starting from initial
conditions that fulfilled the Friedmann constraint, this very
trajectory originates from sections of the phase space that do not
fulfill the constraint, neither do they contain physical value for
all dynamical variables. We also met this peculiar situation in
the de Sitter cosmology case, so it seems there is some universal
behaviour for the exponential Einstein-Gauss-Bonnet models. What
was expected from the qualitative analysis and confirmed from the
numerical integrations is that only at specific cases there exists
a viable and stable equilibrium point where the system may
eventually be attracted to, after relatively large oscillations of
some phase space variables.

Cases may exist -such as when $\lambda < 3\mu$- that the
equilibrium point will exist but it will be neither viable, nor
stable, hence solutions of the model are repelled far away from
it. Furthermore, two important comments are in order, which we
derived from our numerical investigation, which are the following:
\begin{itemize}
\item The convergence to the equilibrium point, if it is stable,
occurs relatively fast, within $20$ $e$-folds. The divergence, on
the other hand, if the equilibrium point is not stable, is even
faster, as some dynamical variables reach extremely large values
within $5$ $e$-folds. \item Even when the initial values of
variables $x_{1}$, $x_{2}$ and $x_{3}$ along the equilibrium they
reach, fulfill the Friedmann constraint, the same is not necessary
throughout the whole trajectory; taking into account the evolution
of $x_{4}$ and $x_{5}$ may solve this paradox.
\end{itemize}
In this case, we should again note, $x_{4}$ is constant while
$x_{5}$ increases as a function of the $e$-foldings number,
diverging from its equilibrium value $x_{5}^{*} = 0$. This should
be attributed either to the behaviour of the Hubble rate, or to
the fact that baryonic matter, especially cold dark matter, is
known to evolve and form structures during the radiation-dominated
phase of the Universe.

\subsection{Matter-dominated Universe: $m = -\dfrac{3}{2}$}

The same behaviour holds true, more or less, in the case the
Universe is dominated by matter. Setting $m = -\dfrac{3}{2}$, then
Eqs. (\ref{eq:firstODE}), (\ref{eq:secondODE}) and
(\ref{eq:thirdODE}) reduce to
\begin{align}
x_{1}^{\prime} &= - \dfrac{3}{2}x_{1} + \dfrac{\sqrt{6}}{2}\lambda x_{2}^2 +2\sqrt{6} x_{3} \; \, , \\
x_{2}^{\prime} &= -x_{2} \Bigg( \dfrac{\sqrt{6}}{2} \lambda x_{1} -\dfrac{3}{2} \Bigg) \;\; \text{and} \\
x_{3}^{\prime} &= x_{3} \Bigg( \dfrac{\sqrt{6}}{2}\mu x_{1}
-\dfrac{3}{2} \Bigg) \, .
\end{align}
Again, these equations exhibit all four equilibrium points we have
mentioned. Point $P_{0}$ possesses two stable and one unstable
manifolds; point $P_{1}$ possesses one stable and one unstable
manifold, while the third, tangent to $\vec{e}_{2} = (0,1,0)$,
shifts from stable to unstable as $\lambda = 3\mu$; finally,
points $P_{2}$ and $P_{3}$ yield two complex eigenvalues with
negative real parts (so attraction with oscillations) and a third
real eigenvalue that shifts from negative to positive as $\lambda
= 3\mu$. The cases of $\lambda$ and $\mu$ used prior, will be used
here as well. Furthermore, in the last case $\lambda = \sqrt{2m(2m
- 3)} = 3\sqrt{2}$ is chosen as to ensure the viability of point
$P_{3}$.

\begin{figure}
\includegraphics[width=0.45\linewidth]{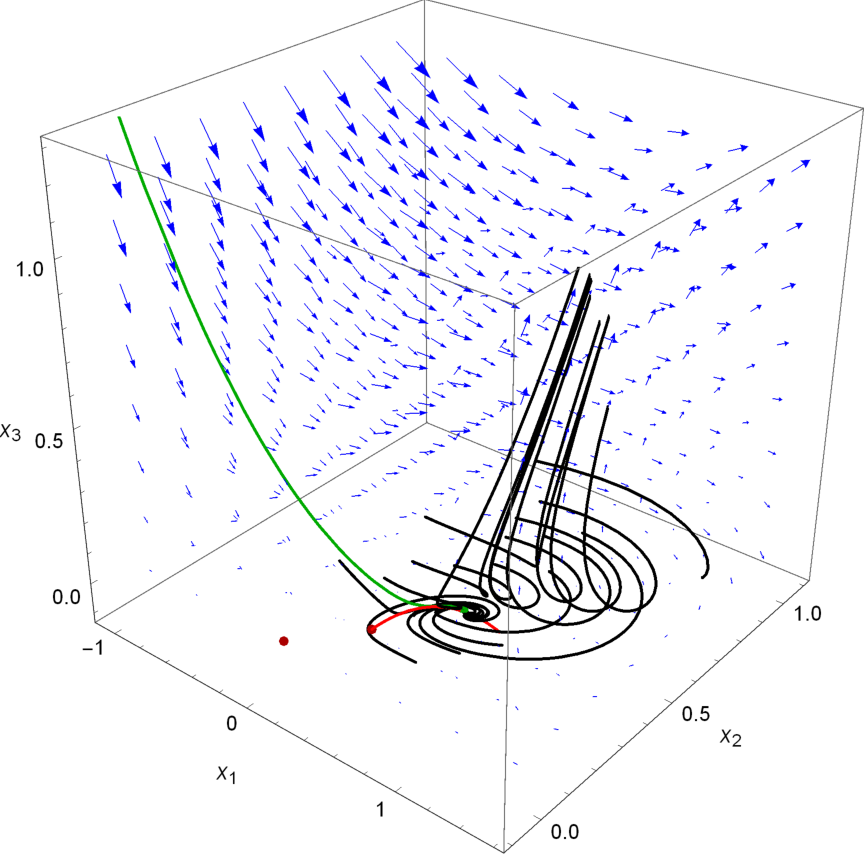}
\includegraphics[width=0.45\linewidth]{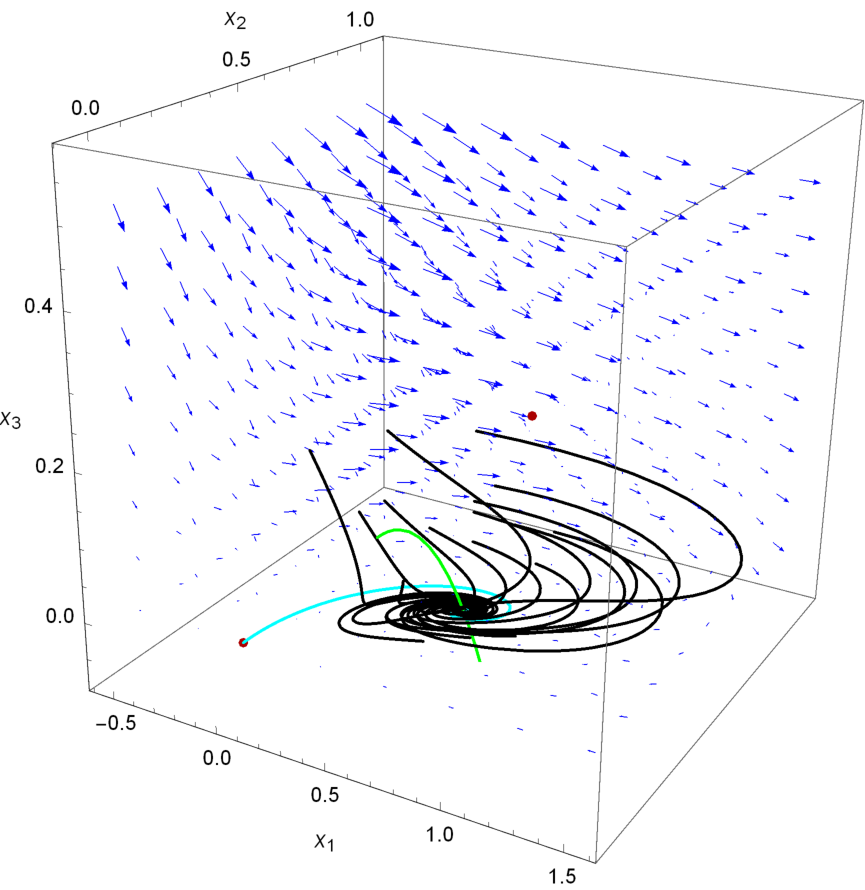}
\includegraphics[width=0.45\linewidth]{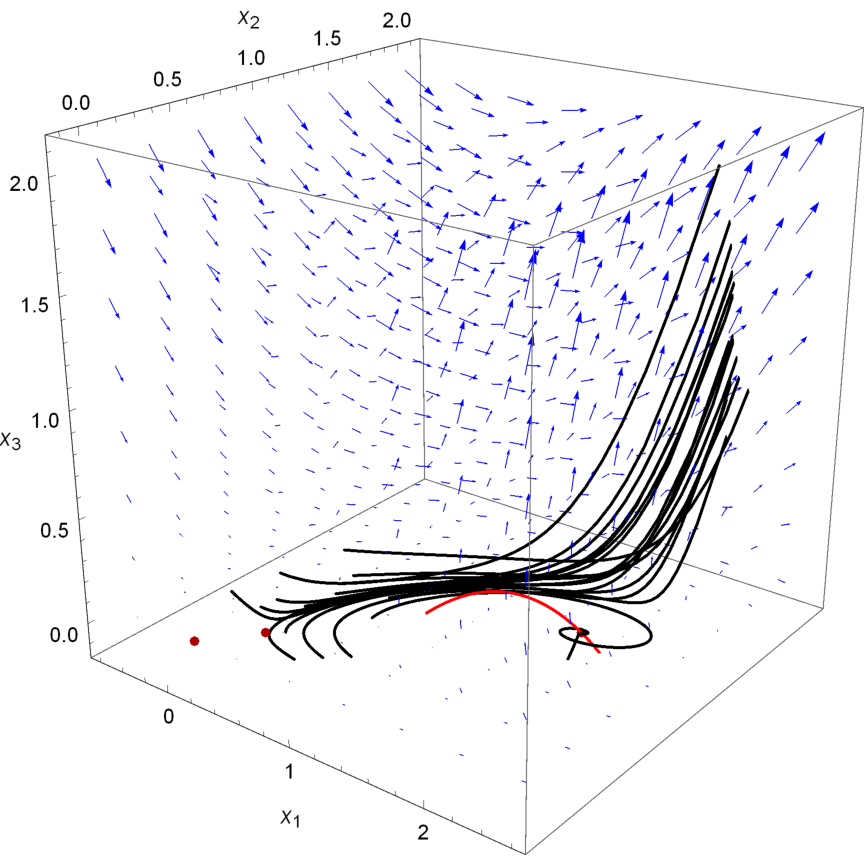}
\includegraphics[width=0.45\linewidth]{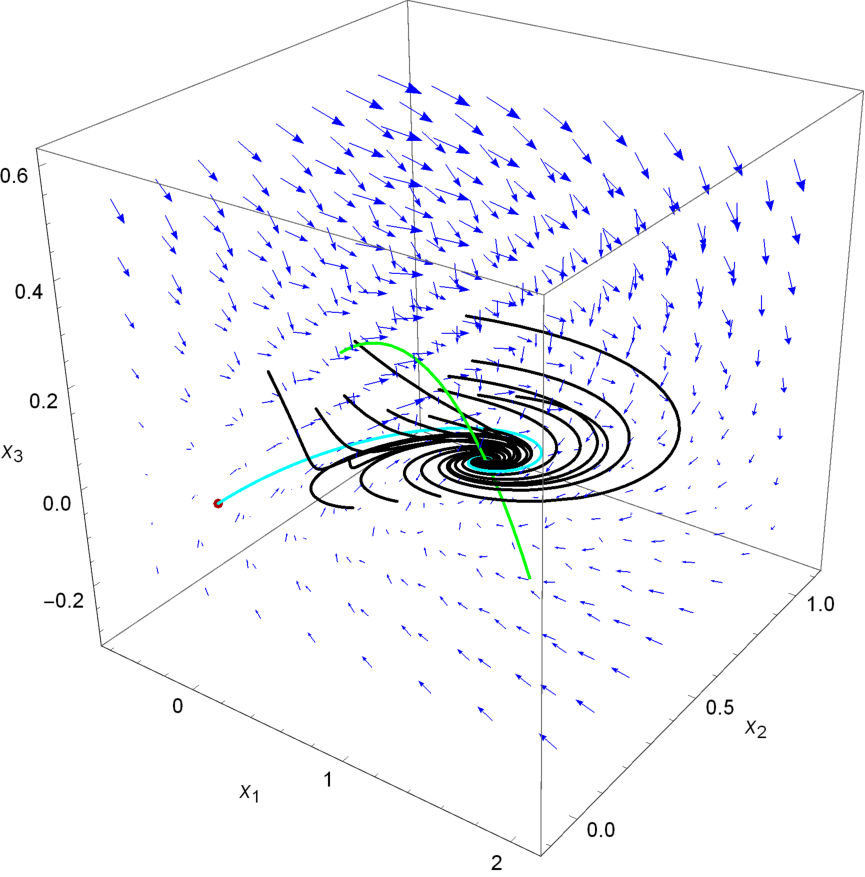}
\caption{The phase space of $x_1$, $x_2$ and $x_3$ for the
matter-dominated Universe, for $\lambda = 3\mu$, $\lambda = 6\mu$,
$\lambda = \mu$ and $\lambda \gg \mu$ (from top-left to
bottom-right). Blue arrows denote the vector field, black lines
denote numerical solutions of the system; the red curve stands for
the unstable manifold and the green curve for the stable manifold,
while the cyan curve depicts the heteroclinic curve from $P_{0}$
to $P_{3}$; the equilibrium points are marked dark red if they are
unviable and dark green if they are viable.}
\label{fig:Matter_Numerics}
\end{figure}

We should notice that the curve $C$ takes the form
\begin{equation}
x_{3} = \dfrac{\frac{9}{2} -\lambda^2 x_{2}^2}{4\lambda}\, ,
\end{equation}
and acts either as stable, centre or unstable manifold for the
system, depending on the relationship between $\lambda$ and $\mu$.
The stable cases are denoted as green curves, whereas the unstable
ones are denoted as red in Fig. \ref{fig:Matter_Numerics}. We also
notice the existence of a heteroclinic curve leading from
equilibrium $P_{0}$ to equilibrium $P_{3}$. This curve is appears
in cyan color in two of the subplots of Fig.
\ref{fig:Matter_Numerics}.

In the case of $\lambda = 3\mu$ (the top-left subplot of Fig.
\ref{fig:Matter_Numerics}, where the manifold is neutral, we
solved the ``inverse problem'' tracing a trajectory that reaches a
physical point in the phase space, with $x_{1} =
\dfrac{3\sqrt{6}}{2\lambda}$ , $x_{2} = \dfrac{1}{34}
\left(\dfrac{189}{\lambda^2} + 2\right)$ and $x_{3} = -\dfrac{4
\lambda^4 -4446 \lambda^2 +35721}{4624 \lambda^3}$. This
trajectory is marked dark green in the first subplot of Fig.
\ref{fig:Matter_Numerics}. Its initiation, unlike all other cases
-marked black- does not fulfill the Friedmann constrains.

In fact, whatever said in the case of the radiation-dominated
Universe holds equally for the matter-dominated. The only
difference concerns the behaviour of discarded variables $x_{4}$
and $x_{5}$. Here, $x_{5}$ is constant while $x_{4}$ decreases
over the $e$-foldings number, eventually reaching its equilibrium
value $x_{4}^{*} = 0$; both evolutions are in accordance to the
Standard Cosmological model, whereas the density of radiation
decreases during the matter-dominated era.

\subsection{Stiff Matter-dominated Universe: $m = -3$}

The case of stiff matter, although more bizarre, seems far more
interesting, since the phase space alters significantly. Setting
$m = -3$ to Eqs. (\ref{eq:firstODE}), (\ref{eq:secondODE}) and
(\ref{eq:thirdODE}), we obtain
\begin{align}
x_{1}^{\prime} &= \dfrac{\sqrt{6}}{2}\lambda x_{2}^2 +8\sqrt{6} x_{3} \; \, , \\
x_{2}^{\prime} &= -x_{2} \Bigg( \dfrac{\sqrt{6}}{2} \lambda x_{1} -3 \Bigg) \;\; \text{and} \\
x_{3}^{\prime} &= x_{3} \Bigg( \dfrac{\sqrt{6}}{2}\mu x_{1} -3
\Bigg) \, .
\end{align}
These equations contain in fact three of the original equilibrium
points, as $P_{2}$ and $P_{3}$ coincide. However, it contains a
peculiar solution, in the form of $x_{2}^{*} = 0$ and $x_{3}^{*} =
0$, which eventually means that the $x_{1}$ axis contains infinite
equilibrium points (including $P_{0}$, $P_{1}$ and $P_{2} =
P_{3}$). It is relatively easy to see that the eigenvector $v_{1}
= \vec{e}_{1} = (1,0,0)$ is always a centre manifold for any of
the infinite equilibrium points on this axis. Furthermore, the
eigenvector $v_{2} = \vec{e}_{2} = (0,2,0)$ is found to be tangent
to a stable manifold for $\lambda > \mu$, a centre manifold for
$\lambda = 3\mu$ and an unstable manifold for $\lambda < 3\mu$. It
is also an attractor for any $x_{1} > \dfrac{3\sqrt{6}}{\lambda}$.
Finally, the third direction, found to be tangent to $v_{3} =
\Big( -\sqrt{\dfrac{2}{3}} \dfrac{4 \lambda}{\lambda -3\mu},0,1
\Big)$ is attractive towards the $x_{1}$ axis for any $\lambda >
3\mu$.

The curve $C$, usually denoting a manifold of the viable
equilibrium point $P_{3}$, though it exists, it is usually reduced
to a simple line. The same holds for the heteroclinic curve
connecting $P_{1}$ with $P_{3}$. Though it still exists, it has
been reduced to a straight line across the $x_{1}$ axis. The phase
space contains least one centre manifold along the $x_{1}$ axis.

\begin{figure}
\includegraphics[width=0.45\linewidth]{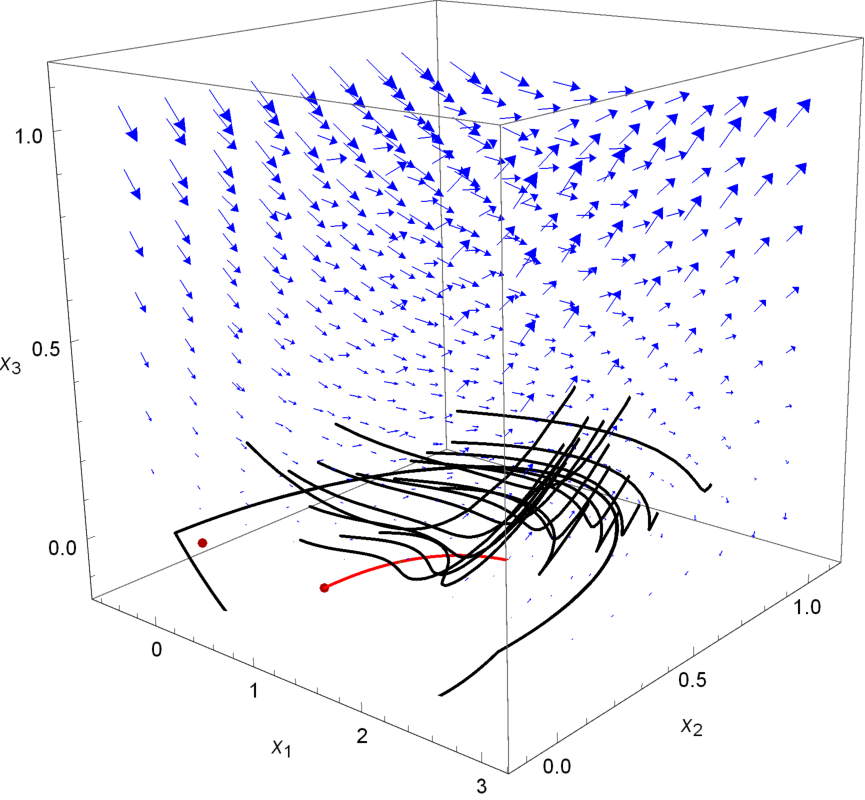}
\includegraphics[width=0.45\linewidth]{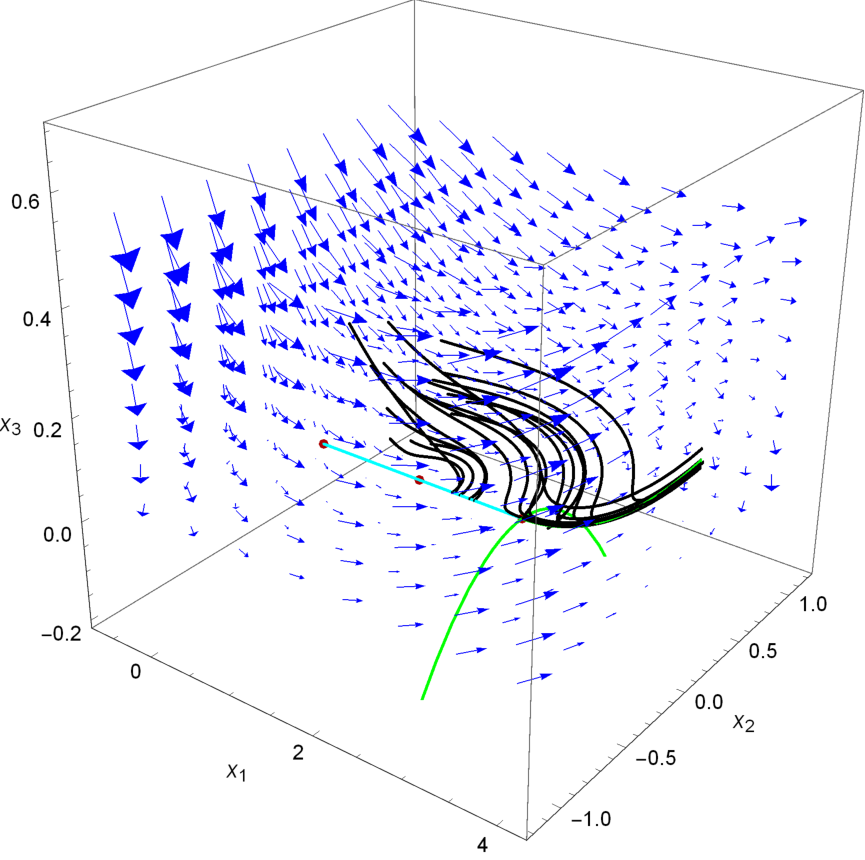}
\includegraphics[width=0.45\linewidth]{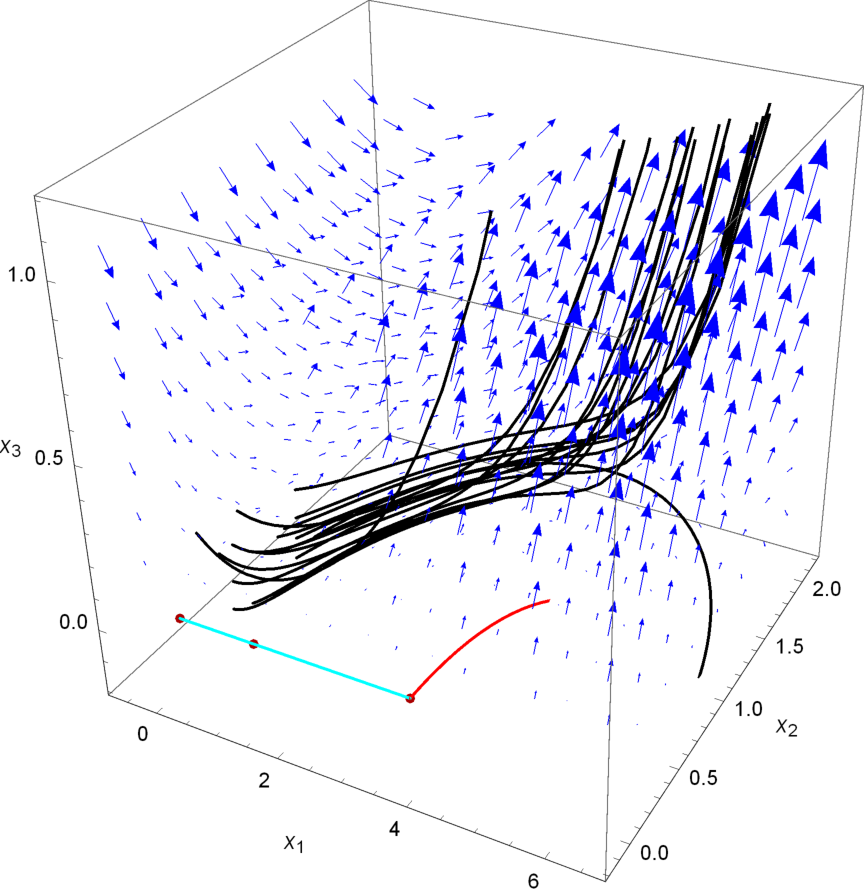}
\includegraphics[width=0.45\linewidth]{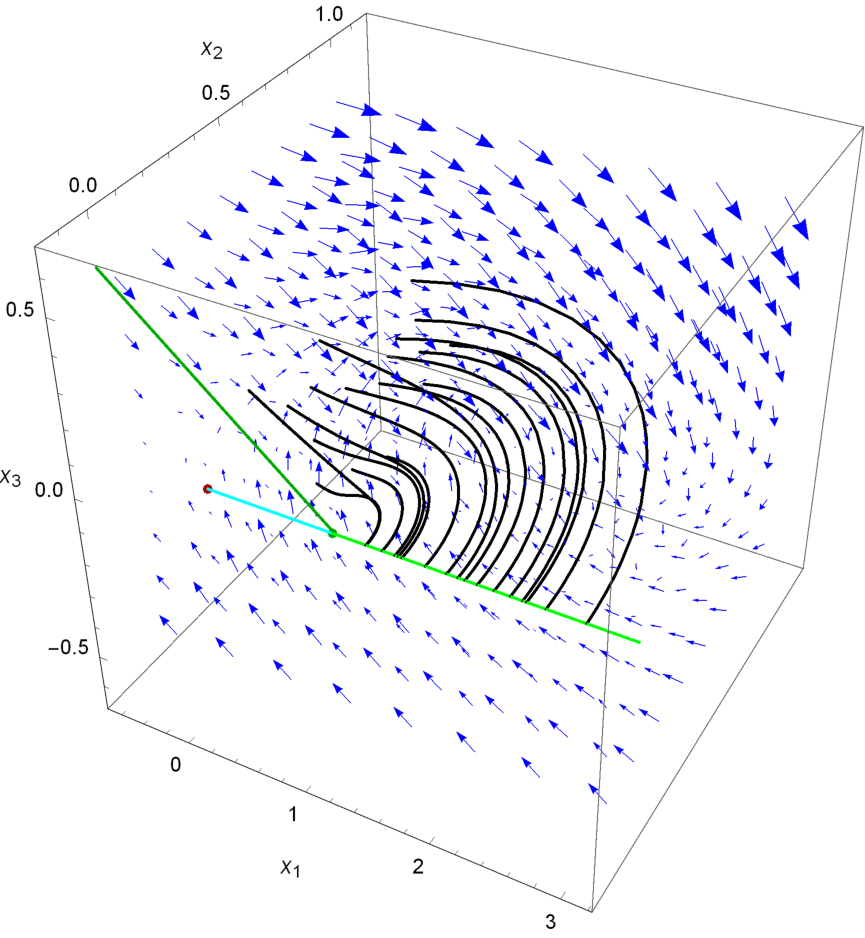}
\caption{The phase space of $x_1$, $x_2$ and $x_3$ for the stiff
matter-dominated Universe, for $\lambda = 3\mu$, $\lambda = 6\mu$,
$\lambda = \mu$ and $\lambda \gg \mu$ (from top-left to
bottom-right). Blue arrows denote the vector field, black lines
denote numerical solutions of the system; the red curve stands for
the unstable manifold and the green curve for the stable manifold,
while the cyan curve depicts the heteroclinic curve from $P_{0}$
to $P_{3}$; the equilibrium points are marked dark red if they are
unviable and dark green if they are viable.}
\label{fig:StiffMatter_Numerics}
\end{figure}
We should especially focus on the case where $\lambda =
\sqrt{2m(2m - 3)} = 3\sqrt{6}$ -depicted in the fourth subplot of
Fig. \ref{fig:StiffMatter_Numerics}- that ensures the viability of
equilibrium $P_{2} = P_{3}$. Here, trajectories begin an
oscillatory motion around equilibrium $P_{2} = P_{3}$, which
eventually leads them to a halt as they meet the horizontal axis
for some $x_{1} > \dfrac{3\sqrt{6}}{\lambda}$, captured by the
centre manifold. Hence, unlike all four previous models examined,
in the case of a stiff matter-dominated Universe, there is no
condition to ensure that the viable equilibrium is also stable and
generally reached by the solutions of the system. Solving the
``inverse problem'', we manage to trace a trajectory leading to
the equilibrium $P_{2} = P_{3}$, however, as with other such case,
it does not originate from a physically meaningful section of the
phase space, neither does it fulfill the Friedmann constraint all
the way through from its beginning -the trajectories obtained by
the ``inverse problem'' are painted dark green in the first and
the fourth subplot of Fig. \ref{fig:StiffMatter_Numerics}.

Once more, both $x_{4}$ and $x_{5}$ increase over the $e$-foldings
number, diverging from the equilibrium $x_{4}^{*} = x_{5}^{*} =
0$. This behaviour is hard to explain, though it probably relies
on the behaviour of the Hubble rate, or the fact that stiff matter
dominates the Universe, rendering both dust and radiation
unstable.


\section{Analysis of the Results and Concluding Remarks}

Let us now elaborate on the results and discuss the outcomes of
the analysis performed in the previous sections. We have seen that
the dynamical system contains up to four equilibrium points, only
one of which is viable and physically meaningful under specific
conditions. It also contains at least one invariant submanifold
that may shift from unstable to stable, either repelling or
attracting solutions from/to the viable equilibrium point, or in
specific cases turn to centre manifold, attracting all solutions
on it. Finally, it contains a heteroclinic trajectory leading from
an unviable equilibrium to the viable one. Our numerical analysis
also showed that the equilibrium is reached relatively fast
(within $20$ to $30$ $e$-folds, for all examined cases), while
achieving a viable equilibrium is not at all secured, even if the
initial conditions fulfill the Friedmann constraint for the
reduced phase space of $x_{1}$, $x_{2}$ and $x_{3}$. In sharp
contrast, it is shown that the viable equilibrium may be reached
by trajectories that do not fulfill the Friedmann constraint when
$x_{4} = x_{5} = 0$ is imposed and contain non-physical values for
at least one dynamical variable, namely a negative value for
$x_{2}$ or $x_{3}$.

Combining these elements with the theory developed in Section II
and the definitions of the dynamical variables in Eq.
(\ref{eq:PhaseVar}), we may provide a clue as to the viability of
the original theory, at least in the specific cases of FRW
cosmologies studied here.

First of all, we should make clear that the viable equilibrium
$P_{3}$ is proved to be stable for most of the cases, provided
that $\lambda = \sqrt{2m(2m - 3)}$ (condition of viability) and
$\lambda > 3\mu$ (condition of stability). Hence it may attract
solutions from the whole phase space. Essentially, the
cosmological model under examination may be attracted on this
stationary state after evolving for some time, due to the presence
of the potential and the Gauss-Bonnet term. As for the meaning of
this equilibrium, one should state that both the behaviour of
$x_{1}$ and $x_{2}$ seems normal. They denote,
\begin{equation*}
\dfrac{1}{2}\dot{\phi}^2 = \dfrac{18}{\lambda^2} \Big(
\dfrac{\dot{H}}{H} \Big)^2 = \dfrac{18}{\lambda^2 t^2} \;\;\;
\text{and} \;\;\; V(\phi) = \dfrac{6}{\lambda^2} \Bigg( \Big(
\dfrac{\dot{H}}{H} \Big)^2 -3\dfrac{\dot{H}}{H} \Bigg) =
\dfrac{6}{\lambda^2} \Big( \dfrac{1}{t^2}-\dfrac{3}{t^2} \Big) \,
,
\end{equation*}
in other words, that the kinetic term and the potential of the
scalar field must evolve according to time as the evolution of the
system ceases; their evolution is proved to be rather trivial,
since both tend to zero, for a large time-scale. Yet, the value of
$x_{3}$ comes quite as peculiar, since it demands that the
coupling function turns constant. This appears reasonable at
first, since it states that at equilibrium, the Gauss-Bonnet term
is decoupled from the scalar field, or even nullified, however, it
turns out to be a non-physical argument. To achieve this, we need
either $h_{0} = 0$ or $\mu = 0$, both of which cannot occur. On
the one hand, setting $\mu = 0$ leads to the emergence of poles on
the phase space, on the other hand, either $h_{0} = 0$ or $\mu =
0$ means that the coupling function should be zero from the
beginning, hence the scalar field should be decoupled from the
Gauss-Bonnet term. This immediately sets us off with respect to
our original theoretical streamline, since we no longer deal with
an Einstein-Gauss-Bonnet model.

The only physical explanations for it would be that either the
assumption of an exponential coupling function was a rather
misleading one, or that the quadratic curvature terms should
vanish at some point, leading to the nullifying of the coupling,
without the necessity of $\mu$ turning to zero. Both of these
explanations, however, fall beyond the reach of the examined
models.

A second interesting case is that of the centre manifold appearing
at $m = 0$ or $\lambda = 3\mu$, in the form of the curve $C$. In
this case,
\begin{equation*}
\dfrac{1}{2}\dot{\phi}^2 = \dfrac{18}{\lambda^2} \Big(
\dfrac{\dot{H}}{H} \Big)^2 \; \, , \;\;\; V(\phi) = \dfrac{18
\dot{H}^2 \big( \dot{H} + 5 H^2 \big) + H^4 \big( \dot{H} + H^2
\big) \lambda^2}{H^2 \big( 5\dot{H} - H^2 \big) \lambda^2} \;\;\;
\text{and} \;\;\; h(\phi) = \dfrac{4 \dot{H}^2 -6\dot{H} H^2 - H^4
\lambda}{8\lambda\mu \big(5 \dot{H} - H^2 \big)} \, ,
\end{equation*}
so that the kinetic term, the potential and the coupling function
end up to some non-zero values, all of them being physical. This
may occur in the following two cases:
\begin{itemize}
\item If $m = 0$, so that $H = H_{0}$ and $\dot{H} = 0$, meaning
that
\begin{equation*}
\dfrac{1}{2} \dot{\phi}^2 = 0 \; \, , \;\;\; V(\phi) = -H_{0}^2
\;\;\; \text{and} \;\;\; h(\phi) = \dfrac{H_{0}^2}{8}
\dfrac{\lambda}{\mu} \, .
\end{equation*}
In this case, the kinetic term vanishes, so the evolution of the
scalar field ceases, but neither the potential nor the coupling to
the Gauss-Bonnet term vanish. Essentially, this may occur during
the early- or late-time accelerating expansion of the Universe,
where the scalar field and quadratic curvature terms may indeed
give rise to the specific behaviour. If it is so, then when the
system is at the final attractor, the Hubble rate remains
constant, so the expansion remains accelerating. Furthermore,
neither the Gauss-Bonnet term, nor the scalar field may completely
vanish, if the accelerating expansion does not come to a halt. The
scalar field turns constant\footnote{Given that $\dot{\phi}^2 =
\dfrac{36}{\lambda^2 t^2}$, then $\phi = \phi_0 \pm
\dfrac{6}{\lambda} \ln t$, that tends very slowly to infinity; if
however we impose that $\dot{\phi} \simeq 0$ in large time-scale,
then $\phi \simeq \phi_0$.} and -along with the Gauss-Bonnet term,
or even on its own- acts as a Cosmological constant.

\item If $\lambda = 3\mu$. In this case, $\lambda$ may be
arbitrary, since the viability of the equilibrium does not depend
on it, however, it must be triple in magnitude in comparison to
$\mu$. This equality means that, given an increasing value of the
scalar field, the potential must decrease in a triple rate to that
of the increase of the coupling function -and the opposite, for a
decreasing scalar field. In other words, the potential must vanish
prior to the coupling of the scalar field to the quadratic
curvature terms. Consequently, quadratic curvature may survive
outside the inflationary era and during the radiation-dominated
and matter-dominated eras, where it should not be active. This
mechanism, however, is capable of retaining the quadratic
curvature terms long after the inflation and re-triggering them at
late-times, so as to give the observed late-time accelerating
expansion.
\end{itemize}

These may sound a good arguments in support of the
Einstein-Gauss-Bonnet theory, providing that an inherent mechanism
exists that would nullify the quadratic curvature coupling to the
scalar field during radiation- and matter-dominated eras and
re-trigger it during the late-time acceleration. In this manner,
both early- and late-time dynamics of the Universe could be
derived from the same model. However, the analysis of the
radiation- and matter-dominated eras, especially the stability of
equilibrium point $P_{3}$, complicate this estimation. As we
demonstrate afterwards the survival of the quadratic curvature
coupling to the scalar field is seriously questioned for the
classical eras.

Third, concerning the specific values of $\lambda$ and $\mu$ in
all cases we  must have $m < 0$, so that equilibrium point $P_{3}$
is both stable and physical. it turns out that $\lambda$ must
track the Hubble rate, through the condition $\lambda =
\sqrt{2m(2m-3)}$ to retain the viability of the equilibrium, which
eventually means that the potential must be somehow interacting
with the expansion of the $3-d$ space. Furthermore, to ensure the
stability of the equilibrium, another condition is imposed,
$\lambda > 3\mu$, which eventually means that $\mu$ is a free
parameter as long as it is smaller enough than $\lambda$. In this
way, we ensure that the coupling function increases less rapidly
than the potential, given an increasing scalar field, and the
quadratic curvature terms may remain coupled to the scalar field
after the early stages of the Universe throughout its whole
evolution, only to be re-triggered in the late-time.

Fourth, a heteroclinic trajectory exists in the phase space,
leading from the unviable equilibrium point $P_{0}$ to the viable
equilibrium point $P_{3}$, retaining the value of variable $x_{3}$
equal to zero. This might be the sole most important trajectory in
the phase space, despite the fact that it leads to an \textit{ab
initio} nihilism of the Gauss-Bonnet term and/or its coupling to
the scalar field. In fact, setting $x_{3} = 0$ and remaining on
the $x_{1}-x_{2}$ plane, we are dealing with the simple form of a
minimally coupled scalar-tensor theory. This plane is perhaps more
important in cases where $m \neq 0$ -the case where the two
equilibrium points are distinct and the heteroclinic curve
actually exists. In such cases, we may assume that the quadratic
curvature terms have vanished, or equivalently that the scalar
field has been decoupled from the Gauss-Bonnet term. Hence, $\mu$
can be chosen to be zero, while $\lambda \neq 0$. In this case,
the phase space would be further reduced on the $x_{1}-x_{2}$
plane, where both the unviable equilibrium $P_{0}$ and the viable
equilibrium $P_{3}$ lie, along with the heteroclinic curve among
them. Subsequently, all we should discuss about would concern the
simplest form of a scalar-tensor theory, maintaining the scalar
field and its potential throughout the radiation-dominated and
matter-dominated eras, so that it would re-emerge in the late-time
era, and cause the accelerating expansion. In such a scenario, the
transcendence from the early Universe (de Sitter case) to the
later stages of evolution (radiation-dominated and
matter-dominated Universe) would demand a double bifurcation.
Firstly, $m$ should turn from zero to negative, and secondly,
$\mu$ should turn from positive to zero. If so, then an early-time
accelerating expansion governed by both the scalar field and the
Gauss-Bonnet term, would lead to standard cosmologies with a
latent scalar field, and eventually to a late-time accelerating
expansion, governed solely by the scalar field.\footnote{The
latter indicates a de Sitter expansion ($m = 0$) for $\mu = 0$, in
other words, to a de Sitter evolution on the $x_{1}$-$x_{2}$ plane
-a case that is unlikely, yet mathematically solid.} The existence
of the heteroclinic trajectory, repelling solutions from $P_{0}$
and attracting them to $P_{3}$, is vital for all models where $m <
0$, since it denotes the slow emergence of the scalar field. As
the dynamical variables rise from zero to $x_{1}^{*} =
-\sqrt{\dfrac{6m^2}{2m^2 - 6m}}$ and $x_{2}^{*} =
\sqrt{-\dfrac{m+3}{m-3}}$, the kinetic term and the potential of
the scalar field turn from zero to non-zero. At the same time, the
non-accounted-for variables $x_{4}$ and $x_{5}$ evolve according
to Eqs. (\ref{eq:x4x5Sol}), namely $x_{4}$ increasing and $x_{5}$
remaining constant during the radiation-dominated era, or $x_{4}$
remaining constant and $x_{5}$ decreasing towards $x_{5}^{*} = 0$
during the matter-dominated era.

\begin{figure}
\includegraphics[width=0.6\linewidth]{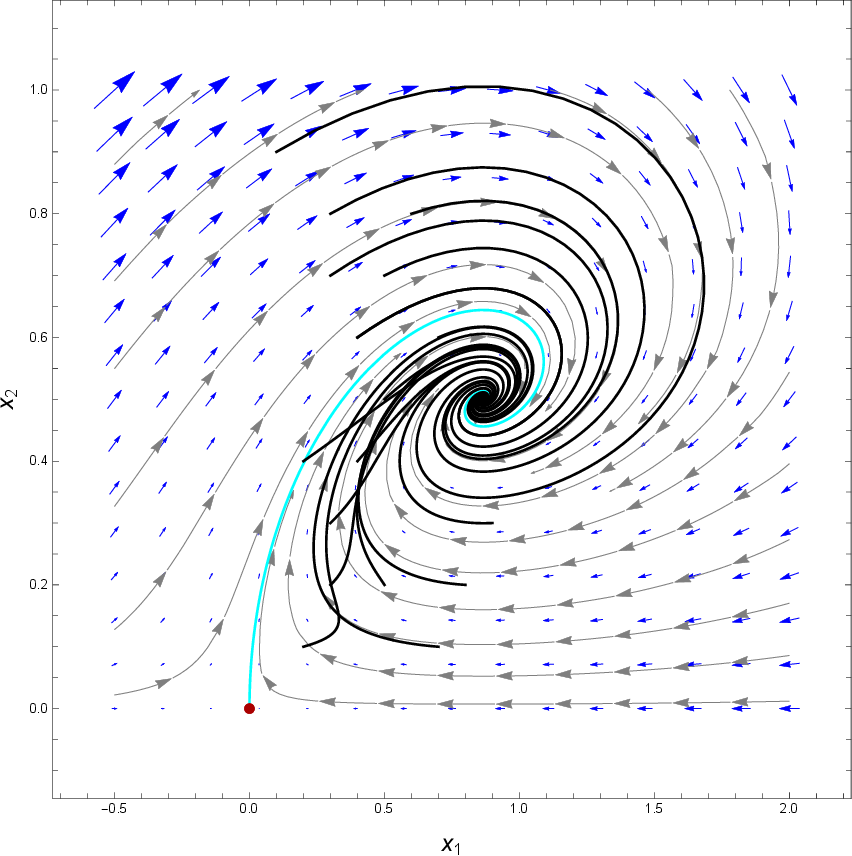}
\caption{The projection of the phase space on the $x_{1}$-$x_{2}$
plane (for $x_{3} = 0$). Blue arrows denote the vector field,
while grey streamlines denote possible evolutions on the plane;
the equilibrium points $P_{1}$ and $P_{2}$ are the crimson and
green spots, while the cyan curve connecting them stands for the
heteroclinic trajectory. The black curves are projections of the
numerical solutions of the complete system; notice that after a
brief time, they fall on the plane and follow its streamlines.}
\end{figure}

Last, but not least, we should also refer to what appears to be a
problem of the non-fulfillment of the Friedmann constraint
throughout the whole evolution of the dynamical variables from
their initial state, until the attainment of an equilibrium
-either $P_{3}$ or the curve $C$. This is, in fact, a very
interesting point, since it allows us to consider two following
two facts that have not been clarified so far, one of them being
the separate evolution of $x_{4}$ and $x_{5}$, the other the
question of the flatness of the Universe.
\begin{itemize}
\item What seems as logical at first, is that our assumption of
the flatness of the Universe was wrong and perhaps a
non-prerequisite. This contradicts both observations confirming
the flatness of the Universe present time. However, imposing
$x_{4} = x_{5} = 0$, hence a vacuum Universe, in the case of
radiation- and matter-dominated eras ($m = -2$ and $m = -3/2$
respectively), where matter fields could be mimicked by the
modification of gravity, as it is accustomed in the literature,
leads to the paradox of trajectories that are unviable at their
initial state, or during their evolution, but converge to a viable
equilibrium point, namely $P_{3}$. It is furthermore proved that
all these trajectories surpass a stage for which $x_{1}^2 +
x_{2}^2 + 8\sqrt{6} x_{1} x_{3} < 1$, namely the Universe, that
eventually becomes flat, needs not to be flat throughout its full
history, but can realize positive spatial curvature that decreases
to zero as the scalar field and the matter fields emerge and
evolve to finally reach equilibrium $P_{3}$. This is a generally
unacceptable result in relativistic cosmologies, that should be
avoided in modified theories, as well, probing for a different
resolution of the paradox.

\item What we actually assumed to derive this paradox was a
Universe empty of matter fields, a case usually considered yet
easy to discharge as unrealistic. Considering ourselves free to
omit variables $x_{4}$ and $x_{5}$, due to the separability and
integrability of Eqs. (\ref{eq:fourthODE}) and
(\ref{eq:fifthODE}), we practically fell pray to this paradox.
Such an omission is mathematically justified, due to the form of
our dynamical system, yet it is not necessarily justified from a
physical standpoint, especially when considering the Friedmann
constraint. Even if for some equilibrium point, where $x_{4}^{*} =
x_{5}^{*} = 0$, the Friedmann constraint is fulfilled, it does not
mean that the constraint will be fulfilled for all trajectories
attaining the equilibrium -hence for all evolutions realizing this
Universe- without taking into account the non-zero values of
$x_{4}$ and $x_{5}$, prior to their nihilism. Eventually, the
non-fulfillment of the reduced Friedmann constraint for
trajectories that reach a viable equilibrium point can be resolved
if we take into account the evolution of $x_{4}$ and $x_{5}$ as
they decrease towards their equilibrium values. Notably, all
trajectories that yield $x_{1}^2 + x_{2}^2 + 8\sqrt{6} x_{1} x_{3}
< 1$ could easily yield $x_{1}^2 + x_{2}^2 + 8\sqrt{6} x_{1} x_{3}
+ x_{4}^2 + x_{5}^2 = 1$ if Eqs. (\ref{eq:fourthODE}) and
(\ref{eq:fifthODE}) were taken into account. It is, in fact, the
combined behaviour of $x_{1}$, $x_{2}$, $x_{4}$ and $x_{5}$ that
may fulfill the Friedmann constraint and ensure physical solutions
for the reduced scalar-tensor theory.

\end{itemize}

To conclude, we may state that the phase space analysis of the
Einstein-Gauss-Bonnet cosmological models revealed a great number
of interesting facts about the theory overall and its ability to
describe the actual evolution of the Universe. It is sound to
assume that it offers a viable and stable equilibrium point, only
if certain conditions are fulfilled for the free parameters of the
theory. Furthermore, while the de Sitter case gives rise to a
viable inflationary scenario, the non-fulfillment of the Friedmann
constraint results to the necessity of matter fields or of an open
Universe during the early and very early Universe; of these two,
the former sounds as the more reasonable choice, deeming unsafe to
treat such cosmological models as a vacuum Universe, even when the
Gauss-Bonnet invariant can mimic the effects of matter fields.
Finally, the transition from the de Sitter inflationary phase to
the later stages of evolution, such as the radiation-dominated and
the matter-dominated Universe, should be followed by a second
transition that would decouple the Gauss-Bonnet term from the
scalar field and essentially nullify the quadratic curvature
terms. In this case, the Einstein-Gauss-Bonnet theory seems more
like a predecessor to the scalar-tensor theory that may describe
the later phase of the Universe, up to the late-time accelerating
expansion. In fact, it is a conclusion of our analysis, that the
scalar field turns constant for large time-scales, if the
equilibrium point $P_{3}$ is attained in radiation-dominated and
matter-dominated eras, justifying for the transition to the
late-time phase of accelerating expansion.

The case of stiff matter dominating the Universe is highly
peculiar and degenerate and can be considered as highly
improbable, as in the Standard Cosmological model. The non-minimal
coupling of the scalar field to the Ricci scalar curvature should
also be taken into consideration, but that is rather left for a
future work. Another element in need of further study is the phase
transition and decoupling of the scalar field from the quadratic
curvature, an element that might be shredded with more light in
the view of extensions of Einstein-Gauss-Bonnet models.

\section*{Acknowledgments}

This work is supported by the DAAD program
``Hochschulpartnerschaften mit Griechenland 2016'' (Projekt
57340132) (V.K.O). V.K.O is indebted to Prof. K. Kokkotas for his
hospitality in the IAAT, University of T\"{u}bingen.


\begin{thebibliography}{99}



\bibitem{Lovelock:1971yv}
  D.~Lovelock,
  J.\ Math.\ Phys.\  {\bf 12} (1971) 498.
  doi:10.1063/1.1665613

\bibitem{Farhoudi:1995rc}
  M.~Farhoudi,
  Gen.\ Rel.\ Grav.\  {\bf 41} (2009) 117
  doi:10.1007/s10714-008-0658-9
  [gr-qc/9510060].






\bibitem{reviews1}
 S.~Nojiri, S.~D.~Odintsov and V.~K.~Oikonomou,
  Phys.\ Rept.\  {\bf 692} (2017) 1
  doi:10.1016/j.physrep.2017.06.001
  [arXiv:1705.11098 [gr-qc]].





\bibitem{reviews2}

S. Nojiri, S.D. Odintsov,
   Phys.\ Rept.\  {\bf 505}, 59 (2011);



\bibitem{reviews3}
S. Nojiri, S.D. Odintsov,
  eConf {\bf C0602061}, 06 (2006)
  [Int.\ J.\ Geom.\ Meth.\ Mod.\ Phys.\  {\bf 4}, 115 (2007)].
 [arXiv:hep-th/0601213];


   \bibitem{reviews4}
 S. Capozziello, M. De Laurentis,
   Phys.\ Rept.\  {\bf 509}, 167 (2011)
   [arXiv:1108.6266 [gr-qc]].
 V.~Faraoni and S.~Capozziello,
  Fundam.\ Theor.\ Phys.\  {\bf 170} (2010).
  doi:10.1007/978-94-007-0165-6



\bibitem{reviews5}

A.~de la Cruz-Dombriz and D.~Saez-Gomez,
  Entropy {\bf 14} (2012) 1717
  doi:10.3390/e14091717
  [arXiv:1207.2663 [gr-qc]].

\bibitem{reviews6}

G.~J.~Olmo,
  Int.\ J.\ Mod.\ Phys.\ D {\bf 20} (2011) 413
  doi:10.1142/S0218271811018925
  [arXiv:1101.3864 [gr-qc]].





\bibitem{Chingangbam:2007yt}
  R.~Chingangbam, M.~Sami, P.~V.~Tretyakov and A.~V.~Toporensky,
  Phys.\ Lett.\ B {\bf 661} (2008) 162
  doi:10.1016/j.physletb.2008.01.070
  [arXiv:0711.2122 [hep-th]].




\bibitem{Nojiri:2005am}
  S.~Nojiri, S.~D.~Odintsov and O.~G.~Gorbunova,
  J.\ Phys.\ A {\bf 39} (2006) 6627
  doi:10.1088/0305-4470/39/21/S62
  [hep-th/0510183].

\bibitem{Sanyal:2006wi}
  A.~K.~Sanyal,
  Phys.\ Lett.\ B {\bf 645} (2007) 1
  doi:10.1016/j.physletb.2006.11.070
  [astro-ph/0608104].

\bibitem{Cognola:2006eg}
  G.~Cognola, E.~Elizalde, S.~Nojiri, S.~D.~Odintsov and S.~Zerbini,
  Phys.\ Rev.\ D {\bf 73} (2006) 084007
  doi:10.1103/PhysRevD.73.084007
  [hep-th/0601008].

\bibitem{Li:2007jm}
  B.~Li, J.~D.~Barrow and D.~F.~Mota,
  Phys.\ Rev.\ D {\bf 76} (2007) 044027
  doi:10.1103/PhysRevD.76.044027
  [arXiv:0705.3795 [gr-qc]].


\bibitem{Escofet:2015gpa}
  A.~Escofet and E.~Elizalde,
  Mod.\ Phys.\ Lett.\ A {\bf 31} (2016) no.17,  1650108
  doi:10.1142/S021773231650108X
  [arXiv:1510.05848 [gr-qc]].

\bibitem{Odintsov:2016hgc}
  S.~D.~Odintsov and V.~K.~Oikonomou,
  Phys.\ Lett.\ B {\bf 760} (2016) 259
  doi:10.1016/j.physletb.2016.06.074
  [arXiv:1607.00545 [gr-qc]].

\bibitem{Oikonomou:2016rrv}
  V.~K.~Oikonomou,
  Astrophys.\ Space Sci.\  {\bf 361} (2016) no.7,  211
  doi:10.1007/s10509-016-2800-6
  [arXiv:1606.02164 [gr-qc]].

\bibitem{Oikonomou:2017ppp}
  V.~K.~Oikonomou,
  Int.\ J.\ Mod.\ Phys.\ D {\bf 27} (2018) no.05,  1850059
  doi:10.1142/S0218271818500591
  [arXiv:1711.03389 [gr-qc]].

\bibitem{vandeBruck:2017voa}
  C.~van de Bruck, K.~Dimopoulos, C.~Longden and C.~Owen,
  arXiv:1707.06839 [astro-ph.CO].

\bibitem{Bamba:2017cjr}
  K.~Bamba, M.~Ilyas, M.~Z.~Bhatti and Z.~Yousaf,
  Gen.\ Rel.\ Grav.\  {\bf 49} (2017) no.8,  112
  doi:10.1007/s10714-017-2276-x
  [arXiv:1707.07386 [gr-qc]].

\bibitem{Fomin:2017vae}
  I.~V.~Fomin and S.~V.~Chervon,
  Grav.\ Cosmol.\  {\bf 23} (2017) no.4,  367
  doi:10.1134/S0202289317040090
  [arXiv:1704.03634 [gr-qc]].

\bibitem{Fomin:2017qta}
  I.~V.~Fomin and S.~V.~Chervon,
  Mod.\ Phys.\ Lett.\ A {\bf 32} (2017) no.25,  1750129
  doi:10.1142/S0217732317501292
  [arXiv:1704.07786 [gr-qc]].

\bibitem{Houndjo:2017jsj}
  M.~J.~S.~Houndjo,
  Eur.\ Phys.\ J.\ C {\bf 77} (2017) no.9,  607
  doi:10.1140/epjc/s10052-017-5171-4
  [arXiv:1706.09315 [gr-qc]].

\bibitem{Saridakis:2017rdo}
  E.~N.~Saridakis,
  Phys.\ Rev.\ D {\bf 97} (2018) no.6,  064035
  doi:10.1103/PhysRevD.97.064035
  [arXiv:1707.09331 [gr-qc]].




\bibitem{Bamba:2009uf}
  K.~Bamba, S.~D.~Odintsov, L.~Sebastiani and S.~Zerbini,
  Eur.\ Phys.\ J.\ C {\bf 67} (2010) 295
  doi:10.1140/epjc/s10052-010-1292-8
  [arXiv:0911.4390 [hep-th]].

\bibitem{vandeBruck:2016xvt}
  C.~van de Bruck, K.~Dimopoulos and C.~Longden,
  Phys.\ Rev.\ D {\bf 94} (2016) no.2,  023506
  doi:10.1103/PhysRevD.94.023506
  [arXiv:1605.06350 [astro-ph.CO]].

\bibitem{Santillan:2017nik}
  O.~P.~Santillan,
  JCAP {\bf 1707} (2017) no.07,  008
  doi:10.1088/1475-7516/2017/07/008
  [arXiv:1703.01713 [gr-qc]].

\bibitem{Bamba:2014zoa}
  K.~Bamba, A.~N.~Makarenko, A.~N.~Myagky and S.~D.~Odintsov,
  JCAP {\bf 1504} (2015) 001
  doi:10.1088/1475-7516/2015/04/001
  [arXiv:1411.3852 [hep-th]].

\bibitem{Mathew:2018rzn}
  J.~Mathew,
  arXiv:1811.06001 [astro-ph.CO].

\bibitem{Makarenko:2012gm}
  A.~N.~Makarenko, V.~V.~Obukhov and I.~V.~Kirnos,
  Astrophys.\ Space Sci.\  {\bf 343} (2013) 481
  doi:10.1007/s10509-012-1240-1
  [arXiv:1201.4742 [gr-qc]].

\bibitem{Kanti:2015pda}
  P.~Kanti, R.~Gannouji and N.~Dadhich,
  Phys.\ Rev.\ D {\bf 92} (2015) no.4,  041302
  doi:10.1103/PhysRevD.92.041302
  [arXiv:1503.01579 [hep-th]].

\bibitem{Kanti:2015dra}
  P.~Kanti, R.~Gannouji and N.~Dadhich,
  Phys.\ Rev.\ D {\bf 92} (2015) no.8,  083524
  doi:10.1103/PhysRevD.92.083524
  [arXiv:1506.04667 [hep-th]].









\bibitem{Canfora:2016umq}
  F.~Canfora, A.~Giacomini, S.~A.~Pavluchenko and A.~Toporensky,
  Grav.\ Cosmol.\  {\bf 24} (2018) no.1,  28
  doi:10.1134/S0202289318010048
  [arXiv:1605.00041 [gr-qc]].

\bibitem{Pavluchenko:2018cmw}
  S.~Pavluchenko,
  Particles {\bf 1} (2018) no.1,  36
  doi:10.3390/particles1010004
  [arXiv:1803.01887 [hep-th]].

\bibitem{Toporensky:2018xpo}
  A.~Toporensky and S.~Pavluchenko,
  EPJ Web Conf.\  {\bf 168} (2018) 02003.
  doi:10.1051/epjconf/201816802003


\bibitem{Shinkai:2017xkx}
  H.~A.~Shinkai and T.~Torii,
  Phys.\ Rev.\ D {\bf 96} (2017) no.4,  044009
  doi:10.1103/PhysRevD.96.044009
  [arXiv:1706.02070 [gr-qc]].


\bibitem{Pavluchenko:2016wvi}
  S.~A.~Pavluchenko,
  Phys.\ Rev.\ D {\bf 94} (2016) no.2,  024046
  doi:10.1103/PhysRevD.94.024046
  [arXiv:1605.01456 [hep-th]].

\bibitem{Pavluchenko:2016hfu}
  S.~A.~Pavluchenko,
  Phys.\ Rev.\ D {\bf 94} (2016) no.8,  084019
  doi:10.1103/PhysRevD.94.084019
  [arXiv:1607.07347 [hep-th]].

\bibitem{Pavluchenko:2017svq}
  S.~A.~Pavluchenko,
  Eur.\ Phys.\ J.\ C {\bf 77} (2017) no.8,  503
  doi:10.1140/epjc/s10052-017-5056-6
  [arXiv:1705.02578 [hep-th]].

\bibitem{Pavluchenko:2018sga}
  S.~A.~Pavluchenko,
  Eur.\ Phys.\ J.\ C {\bf 79} (2019) no.2,  111
  doi:10.1140/epjc/s10052-019-6624-8
  [arXiv:1810.00050 [gr-qc]].

\bibitem{Ivashchuk:2009hi}
  V.~D.~Ivashchuk,
  Grav.\ Cosmol.\  {\bf 16} (2010) 118
  doi:10.1134/S0202289310020040
  [arXiv:0909.5462 [gr-qc]].

\bibitem{Ivashchuk:2016jpe}
  V.~D.~Ivashchuk,
  Eur.\ Phys.\ J.\ C {\bf 76} (2016) no.8,  431
  doi:10.1140/epjc/s10052-016-4284-5
  [arXiv:1607.01244 [hep-th]].


\bibitem{Odintsov:2018zhw}
  S.~D.~Odintsov and V.~K.~Oikonomou,
  Phys.\ Rev.\ D {\bf 98} (2018) no.4,  044039
  doi:10.1103/PhysRevD.98.044039
  [arXiv:1808.05045 [gr-qc]].


\bibitem{Nojiri:2019dwl}
  S.~Nojiri, S.~D.~Odintsov, V.~K.~Oikonomou, N.~Chatzarakis and T.~Paul,
  Eur.\ Phys.\ J.\ C {\bf 79} (2019) no.7,  565
  doi:10.1140/epjc/s10052-019-7080-1
  [arXiv:1907.00403 [gr-qc]].














\bibitem{Glavan:2019inb}
  D.~Glavan and C.~Lin,
  arXiv:1905.03601 [gr-qc].



\bibitem{Bamba:2007ef}
  K.~Bamba, Z.~K.~Guo and N.~Ohta,
  Prog.\ Theor.\ Phys.\  {\bf 118} (2007) 879
  doi:10.1143/PTP.118.879
  [arXiv:0707.4334 [hep-th]].






\bibitem{Koshelev:2013lfm}
  A.~S.~Koshelev,
  Class.\ Quant.\ Grav.\  {\bf 30} (2013) 155001
  doi:10.1088/0264-9381/30/15/155001
  [arXiv:1302.2140 [astro-ph.CO]].

\bibitem{Oikonomou:2015qha}
  V.~K.~Oikonomou,
  Phys.\ Rev.\ D {\bf 92} (2015) no.12,  124027
  doi:10.1103/PhysRevD.92.124027
  [arXiv:1509.05827 [gr-qc]].


\bibitem{Chakraborty:2018scm}
  S.~Chakraborty, T.~Paul and S.~SenGupta,
  Phys.\ Rev.\ D {\bf 98} (2018) no.8,  083539
  doi:10.1103/PhysRevD.98.083539
  [arXiv:1804.03004 [gr-qc]].


\bibitem{Maeda:2006pm}
  H.~Maeda,
  Phys.\ Rev.\ D {\bf 73} (2006) 104004
  doi:10.1103/PhysRevD.73.104004
  [gr-qc/0602109].

\bibitem{Benkel:2016rlz}
  R.~Benkel, T.~P.~Sotiriou and H.~Witek,
  Class.\ Quant.\ Grav.\  {\bf 34} (2017) no.6,  064001
  doi:10.1088/1361-6382/aa5ce7
  [arXiv:1610.09168 [gr-qc]].

\bibitem{Abbas:2018ica}
  G.~Abbas and M.~Tahir,
  Eur.\ Phys.\ J.\ Plus {\bf 133} (2018) no.11,  476
  doi:10.1140/epjp/i2018-12274-8
  [arXiv:1811.00382 [gr-qc]].



\bibitem{Oikonomou:2019muq}
  V.~K.~Oikonomou and N.~Chatzarakis,
  arXiv:1905.01904 [gr-qc].



\bibitem{Oikonomou:2019nmm}
  V.~K.~Oikonomou,
  arXiv:1907.02600 [gr-qc].






\bibitem{Odintsov:2018uaw}
  S.~D.~Odintsov and V.~K.~Oikonomou,
  Phys.\ Rev.\ D {\bf 98} (2018) no.2,  024013
  doi:10.1103/PhysRevD.98.024013
  [arXiv:1806.07295 [gr-qc]].

\bibitem{Odintsov:2018awm}
  S.~D.~Odintsov and V.~K.~Oikonomou,
  Phys.\ Rev.\ D {\bf 97} (2018) no.12,  124042
  doi:10.1103/PhysRevD.97.124042
  [arXiv:1806.01588 [gr-qc]].



\bibitem{Boehmer:2014vea}
  C.~G.~Boehmer and N.~Chan,
  doi:10.1142/9781786341044.0004
  arXiv:1409.5585 [gr-qc].




\bibitem{Bohmer:2010re}
  C.~G.~Boehmer, T.~Harko and S.~V.~Sabau,
  Adv.\ Theor.\ Math.\ Phys.\  {\bf 16} (2012) no.4,  1145
  doi:10.4310/ATMP.2012.v16.n4.a2
  [arXiv:1010.5464 [math-ph]].





\bibitem{Goheer:2007wu}
  N.~Goheer, J.~A.~Leach and P.~K.~S.~Dunsby,
  Class.\ Quant.\ Grav.\  {\bf 24} (2007) 5689
  doi:10.1088/0264-9381/24/22/026
  [arXiv:0710.0814 [gr-qc]].





\bibitem{Leon:2014yua}
  G.~Leon and E.~N.~Saridakis,
  JCAP {\bf 1504} (2015) no.04,  031
  doi:10.1088/1475-7516/2015/04/031
  [arXiv:1501.00488 [gr-qc]].


\bibitem{Guo:2013swa}
  J.~Q.~Guo and A.~V.~Frolov,
  Phys.\ Rev.\ D {\bf 88} (2013) no.12,  124036
  doi:10.1103/PhysRevD.88.124036
  [arXiv:1305.7290 [astro-ph.CO]].


\bibitem{Leon:2010pu}
  G.~Leon and E.~N.~Saridakis,
  Class.\ Quant.\ Grav.\  {\bf 28} (2011) 065008
  doi:10.1088/0264-9381/28/6/065008
  [arXiv:1007.3956 [gr-qc]].


\bibitem{deSouza:2007zpn}
  J.~C.~C.~de Souza and V.~Faraoni,
  Class.\ Quant.\ Grav.\  {\bf 24} (2007) 3637
  doi:10.1088/0264-9381/24/14/006
  [arXiv:0706.1223 [gr-qc]].

\bibitem{Giacomini:2017yuk}
  A.~Giacomini, S.~Jamal, G.~Leon, A.~Paliathanasis and J.~Saavedra,
  Phys.\ Rev.\ D {\bf 95} (2017) no.12,  124060
  doi:10.1103/PhysRevD.95.124060
  [arXiv:1703.05860 [gr-qc]].


\bibitem{Kofinas:2014aka}
  G.~Kofinas, G.~Leon and E.~N.~Saridakis,
  Class.\ Quant.\ Grav.\  {\bf 31} (2014) 175011
  doi:10.1088/0264-9381/31/17/175011
  [arXiv:1404.7100 [gr-qc]].


\bibitem{Leon:2012mt}
  G.~Leon and E.~N.~Saridakis,
  JCAP {\bf 1303} (2013) 025
  doi:10.1088/1475-7516/2013/03/025
  [arXiv:1211.3088 [astro-ph.CO]].



\bibitem{Gonzalez:2006cj}
  T.~Gonzalez, G.~Leon and I.~Quiros,
  Class.\ Quant.\ Grav.\  {\bf 23} (2006) 3165
  doi:10.1088/0264-9381/23/9/025
  [astro-ph/0702227].



\bibitem{Alho:2016gzi}
  A.~Alho, S.~Carloni and C.~Uggla,
  JCAP {\bf 1608} (2016) no.08,  064
  doi:10.1088/1475-7516/2016/08/064
  [arXiv:1607.05715 [gr-qc]].


\bibitem{Biswas:2015cva}
  S.~K.~Biswas and S.~Chakraborty,
  Int.\ J.\ Mod.\ Phys.\ D {\bf 24} (2015) no.07,  1550046
  doi:10.1142/S0218271815500467
  [arXiv:1504.02431 [gr-qc]].


\bibitem{Muller:2014qja}
  D.~Muller, V.~C.~de Andrade, C.~Maia, M.~J.~Reboucas and A.~F.~F.~Teixeira,
  Eur.\ Phys.\ J.\ C {\bf 75} (2015) no.1,  13
  doi:10.1140/epjc/s10052-014-3227-2
  [arXiv:1405.0768 [astro-ph.CO]].





\bibitem{Mirza:2014nfa}
  B.~Mirza and F.~Oboudiat,
  Int.\ J.\ Geom.\ Meth.\ Mod.\ Phys.\  {\bf 13} (2016) no.09,  1650108
  doi:10.1142/S0219887816501085
  [arXiv:1412.6640 [gr-qc]].


\bibitem{Rippl:1995bg}
  S.~Rippl, H.~van Elst, R.~K.~Tavakol and D.~Taylor,
  Gen.\ Rel.\ Grav.\  {\bf 28} (1996) 193
  doi:10.1007/BF02105423
  [gr-qc/9511010].


\bibitem{Ivanov:2011vy}
  M.~M.~Ivanov and A.~V.~Toporensky,
  Grav.\ Cosmol.\  {\bf 18} (2012) 43
  doi:10.1134/S0202289312010100
  [arXiv:1106.5179 [gr-qc]].


\bibitem{Khurshudyan:2016qox}
  M.~Khurshudyan,
  Int.\ J.\ Geom.\ Meth.\ Mod.\ Phys.\  {\bf 14} (2016) no.03,  1750041.
  doi:10.1142/S0219887817500414


\bibitem{Boko:2016mwr}
  R.~D.~Boko, M.~J.~S.~Houndjo and J.~Tossa,
  Int.\ J.\ Mod.\ Phys.\ D {\bf 25} (2016) no.10,  1650098
  doi:10.1142/S021827181650098X
  [arXiv:1605.03404 [gr-qc]].

\bibitem{Odintsov:2017icc}
  S.~D.~Odintsov, V.~K.~Oikonomou and P.~V.~Tretyakov,
  Phys.\ Rev.\ D {\bf 96} (2017) no.4,  044022
  doi:10.1103/PhysRevD.96.044022
  [arXiv:1707.08661 [gr-qc]].


\bibitem{Granda:2017dlx}
  L.~N.~Granda and D.~F.~Jimenez,
  arXiv:1710.07273 [gr-qc].



\bibitem{Landim:2016gpz}
  F.~F.~Bernardi and R.~G.~Landim,
  Eur.\ Phys.\ J.\ C {\bf 77} (2017) no.5,  290
  doi:10.1140/epjc/s10052-017-4858-x
  [arXiv:1607.03506 [gr-qc]].


\bibitem{Landim:2015uda}
  R.~C.~G.~Landim,
  Eur.\ Phys.\ J.\ C {\bf 76} (2016) no.1,  31
  doi:10.1140/epjc/s10052-016-3894-2
  [arXiv:1507.00902 [gr-qc]].

\bibitem{Landim:2016dxh}
  R.~C.~G.~Landim,
  Eur.\ Phys.\ J.\ C {\bf 76} (2016) no.9,  480
  doi:10.1140/epjc/s10052-016-4328-x
  [arXiv:1605.03550 [hep-th]].

\bibitem{Bari:2018edl}
  P.~Bari, K.~Bhattacharya and S.~Chakraborty,
  arXiv:1805.06673 [gr-qc].

\bibitem{Chakraborty:2018bxh}
  S.~Chakraborty,
  arXiv:1805.03237 [gr-qc].


\bibitem{Ganiou:2018dta}
  M.~G.~Ganiou, P.~H.~Logbo, M.~J.~S.~Houndjo and J.~Tossa,
  arXiv:1805.00332 [gr-qc].


\bibitem{Shah:2018qkh}
  P.~Shah, G.~C.~Samanta and S.~Capozziello,
  arXiv:1803.09247 [gr-qc].





\bibitem{Odintsov:2017tbc}
  S.~D.~Odintsov and V.~K.~Oikonomou,
  Phys.\ Rev.\ D {\bf 96} (2017) no.10,  104049
  doi:10.1103/PhysRevD.96.104049
  [arXiv:1711.02230 [gr-qc]].


\bibitem{Dutta:2017fjw}
  J.~Dutta, W.~Khyllep, E.~N.~Saridakis, N.~Tamanini and S.~Vagnozzi,
  JCAP {\bf 1802} (2018) 041
  doi:10.1088/1475-7516/2018/02/041
  [arXiv:1711.07290 [gr-qc]].





\bibitem{Odintsov:2015wwp}
  S.~D.~Odintsov and V.~K.~Oikonomou,
  Phys.\ Rev.\ D {\bf 93} (2016) no.2,  023517
  doi:10.1103/PhysRevD.93.023517
  [arXiv:1511.04559 [gr-qc]].


\bibitem{Kleidis:2018cdx}
  K.~Kleidis and V.~K.~Oikonomou,
  arXiv:1808.04674 [gr-qc].



\bibitem{Oikonomou:2019boy}
  V.~K.~Oikonomou,
  Phys.\ Rev.\ D {\bf 99} (2019) no.10,  104042
  doi:10.1103/PhysRevD.99.104042
  [arXiv:1905.00826 [gr-qc]].

\bibitem{tsuji}
  L.~Amendola and S.~Tsujikawa, ``Dark Energy : Theory and
  Observations,''Cambridge University Press

\end{thebibliography}
\end{document}